\documentclass[11pt]{article} 
\usepackage[hmargin=2.5cm,vmargin=3cm]{geometry}
\normalsize
\usepackage{amsfonts}
\usepackage{amssymb}
\usepackage{amscd}
\usepackage{graphics}
\usepackage[dvips]{graphicx}
\usepackage{epsfig}
\usepackage{subfigure}
\usepackage[cmex10]{amsmath}
\usepackage{array}
\usepackage{eqparbox}
\usepackage{hyperref}
\usepackage{fancyhdr}
\usepackage{appendix}
\usepackage{titling}
\usepackage{enumerate}

\DeclareMathAlphabet{\mathbsf}{OT1}{cmss}{bx}{n}
\DeclareMathAlphabet{\mathssf}{OT1}{cmss}{m}{sl}
\DeclareMathAlphabet{\mathcsf}{OT1}{cmss}{sbc}{n}

\newcommand{\genericRV}[1]{\mathssf{#1}}

\newcommand{\genericRVS}[1]{\mathbsf{#1}}

\newcommand{\n}{\mathrm{n}}
\newcommand{\N}{\mathrm{N}}
\newcommand{\ie}{{\em i.e.}}
\newcommand{\etc}{{\em etc}}
\newcommand{\eg}{{\em e.g.}}

\newcommand{\apriori}{{\em a priori}}
\newcommand{\etal}{\textit{et.\ al}}

\newcommand{\secref}[1]{Section~\ref{#1}}
\newcommand{\figref}[1]{Fig.~\ref{#1}}

\newcommand{\thrmref}[1]{Theorem~\ref{#1}}

\newcommand{\appref}[1]{Appendix~\ref{#1}}

\newcommand{\keywords}[1]{Keywords -}

\makeatletter
\def\blfootnote{\xdef\@thefnmark{}\@footnotetext}
\makeatother

\newtheorem{theorem}{Theorem}[section]

\newcommand{\qed}{\nobreak \ifvmode \relax \else
      \ifdim\lastskip<1.5em \hskip-\lastskip
      \hskip1.5em plus0em minus0.5em \fi \nobreak
      \vrule height0.75em width0.5em depth0.25em\fi}

\def\BibTeX{{\rm B\kern-.05em{\sc i\kern-.025em b}\kern-.08em
    T\kern-.1667em\lower.7ex\hbox{E}\kern-.125emX}}

\usepackage{newlfont}
\date{}
\begin{document}
\title{Capacity Bounds for State-Dependent Broadcast Channels}
\author{K. G. Nagananda, Chandra R. Murthy and Shalinee Kishore\thanks{K. G. Nagananda and Shalinee Kishore are with the Dept. of ECE at Lehigh University, Bethlehem, PA, U.S.A. E-mail: \{\texttt{kgn209,skishore\}@lehigh.edu}; Chandra R. Murthy is with the Dept. of ECE at the Indian Institute of Science, Bangalore, India. E-mail: \texttt{cmurthy@ece.iisc.ernet.in}. Corresponding author: K. G. Nagananda.}}
\setlength{\droptitle}{-1in}
\maketitle

\begin{abstract}
In this paper, we derive information-theoretic performance limits for three classes of two-user state-dependent discrete memoryless broadcast channels, with noncausal side-information at the encoder. The first class of channels comprises a sender broadcasting two independent messages to two non-cooperating receivers; for channels of the second class, each receiver is given the message it need not decode; and the third class comprises channels where the sender is constrained to keep each message confidential from the unintended receiver. We derive inner bounds for all the three classes of channels. For the first and second class of channels, we discuss the rate penalty on the achievable region for having to deal with side-information. For channels of third class, we characterize the rate penalties for having to deal not only with side-information, but also to satisfy confidentiality constraints. We then derive outer bounds, where we present an explicit characterization of sum-rate bounds for the first and third class of channels. For channels of the second class, we show that our outer bounds are within a fixed gap away from the achievable rate region, where the gap is independent of the distribution characterizing this class of channels. The channel models presented in this paper are useful variants of the classical broadcast channel, and provide fundamental building blocks for cellular downlink communications with side-information, such as fading in the wireless medium, interference caused by neighboring nodes in the network, {\etc}. at the encoder; two-way relay communications; and secure wireless broadcasting.
\end{abstract}


%

\section{Introduction}\label{sec:introduction}
The information-theoretic study of broadcast channels (BC) was initiated first by Cover in \cite{Cover1972}. In the classical setting, the BC comprises a sender who wishes to transmit $k$ independent messages to $k$ noncooperative receivers. The largest known inner bound on the capacity region when $k=2$ was derived by Marton \cite{Marton1979}. Recently, some ideas were discussed in \cite{Gohari2010}, that is conjectured to lead to a larger inner bound. Capacity outer bounds were presented by Sato in \cite{Sato1978} by utilizing the fact that the capacity region of BC depends on the marginal transition probabilities. Nair and El Gamal provided outer bounds for the two-user case \cite{Nair2007}, based on the results of the more capable BC \cite{Gamal1979}. Liang {\etal} generalized the outer bounds of \cite{Nair2007} by deriving the \emph{New-Jersey} outer bound. Some properties of the \emph{New-Jersey} outer bound were exposed in \cite{Nair2010}, where it was shown to be equivalent to the computable UVW-bound with bounded cardinalities of the auxiliary random variables.

Several variants of this classical setting have also received considerable attention. One of the most prominent variants is the state-dependent BC with side-information, where the probability distribution characterizing the channel depends on a state process, and with the channel state made available as side-information at the transmitter, or at the receiver, or at both ends. Capacity inner bounds for the two-user BC with noncausal side-information at the transmitter were derived in \cite{Steinberg2005a}, where Marton's achievability scheme was extended to state-dependent channels. In \cite{Steinberg2005}, inner and outer bounds were derived for the degraded BC with noncausal side-information at the transmitter; the capacity region was derived when side-information was obtained to the encoder in a causal manner. The capacity region for BC with receiver side-information was derived in \cite{Kramer2007}, where a genie provides each receiver with the message it need not decode. To the best of the authors' knowledge, outer bounds for the two-user BC with noncausal side-information at the encoder have not appeared in the literature.

Yet another issue in wireless communications, owing to the broadcast nature of the wireless medium, is related to information security. That is, the broadcast nature of wireless networks facilitates malicious or unauthorized access to confidential data, denial of service attacks, corruption of sensitive data, \etc. An information-theoretic approach to address problems related to security has gained rapid momentum, and is commonly referred to as information-theoretic confidentiality or wireless physical-layer security \cite{Liang2009}. An information-theoretic approach to secure broadcasting was inspired by the pioneering work of Csisz\'{a}r and K\"{o}rner \cite{Csisz'ar1978}, who derived capacity bounds for the two-user BC, when the sender transmits a private message to $\mathrm{receiver}$ $1$ and a common message to both receivers, while keeping the private message confidential from $\mathrm{receiver}$ $2$. Secure broadcasting with a single transmitter and multiple receivers in the presence of an external eavesdropper was considered in \cite{Ekrem2009}, where the secrecy capacity region was obtained for several special classes of channels. In \cite{Liu2008}, capacity bounds were derived for BC where a sender broadcasts two independent messages to two receivers, while keeping each message confidential from the unintended receiver. Capacity results and bounds for Gaussian BC with confidential messages were reported in \cite{Bagherikaram2009} - \nocite{Liu2009}\cite{Ly2010}. The reader is referred to \cite{Liang2009a} for a comprehensive review of physical-layer security in BC. However, to the best of the authors' knowledge, the joint problem of side-information and confidentiality on the BC has not been addressed in the literature.

\subsection{Main contributions}\label{subsec:ourcontribution}
\begin{figure}[t]
\centering
\includegraphics[height=1.5in,width=6.1in]{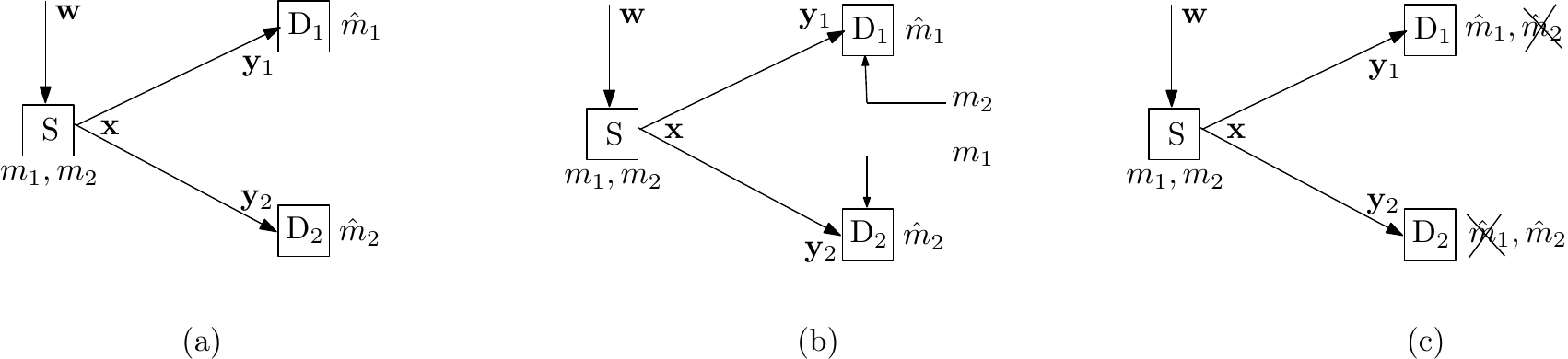}
\caption{State-dependent broadcast channels with side-information at the transmitter: (a) $\mathrm{Class~I}$; (b) $\mathrm{Class~II}$; and (c) $\mathrm{Class~III}$.}\label{fig:bc_sideinfo2}
\end{figure}

In this paper, we aim to provide useful insights into the effect of noncausal side-information at the encoder on $(1)$ the classical two-user BC; $(2)$ the BC with genie- aided receiver side-information; and $(3)$ the BC with confidentiality constraints on the messages. Towards this end, we define three different classes of two-user discrete memoryless BC with noncausal side-information at the encoder. Of particular interest is the $\mathrm{Class~III}$ channels (described below), which provides a fundamental building block to jointly address side-information and confidentiality in BC.
\begin{enumerate}
\item $\mathrm{Class~I}$: A sender broadcasts two independent messages to two non-cooperating receivers (see  \figref{fig:bc_sideinfo2}(a)). We derive an inner bound for this class of channels and characterize the rate penalty for dealing with noncausal side-information at the encoder. We are mainly concerned with outer bounds for this class of channels, where we present an explicit single-letter characterization of the sum-rate bound, along with bounds on single-user rates. An example for $\mathrm{Class~I}$ channels is a base-station transmitting to two mobile receivers, with the base-station having prior knowledge of interference from a transmitter located in its vicinity, {\eg}, through a backhaul network.

\item $\mathrm{Class~II}$: A sender broadcasts two independent messages to two receivers, with each receiver having {\apriori} knowledge of the message it need not decode (see \figref{fig:bc_sideinfo2}(b)). An example of this scenario is full-duplex communications between two nodes, aided by a relay. The relay node broadcasts the messages to the terminals, with each terminal knowing its own message. We devise an achievability scheme to derive an inner bound for this class of channels and show that the achievable rate for each user is in fact the maximum rate achievable for a single-user channel with states known {\apriori} at the encoder. We also derive an outer bound which is within a fixed gap away from the achievable region, where the gap is independent of the distribution characterizing this class of channels.

\item $\mathrm{Class~III}$: A sender broadcasts two independent messages to two receivers, such that each message is kept confidential from the unintended receiver (see \figref{fig:bc_sideinfo2}(c)). To the best of the authors' knowledge, this is the first instance of a study of \emph{simultaneous} impact of side-information and confidentiality constraints on BC. An inner bound for this class of channels is derived employing stochastic encoders to satisfy confidentiality constraints; we characterize the rate penalties for having to deal not only with side-information, but also to satisfy confidentiality constraints. One of the outer bounds is derived by employing a genie, which gives one of the receivers the message it need not decode, while the other receiver computes the equivocation rate treating this message as side-information. We also derive another outer bound, with an explicit characterization of the sum-rate bounds. As an example for this class of channels, we can extend the example considered for $\mathrm{Class~I}$ channels, with the additional constraint of keeping each message confidential from the unintended receiver.
\end{enumerate}

The remainder of the paper is organized as follows. In \secref{sec:systemmodel}, we introduce the notation used and provide a mathematical model for the discrete memoryless version of the channels considered in this paper. In \secref{sec:statementofresults}, we summarize the main results of this paper by describing inner and outer bounds for all the channel models, and provide related discussion. The proofs of the achievability theorems can be found in \secref{sec:achieveproof}, while the proofs of the outer bounds are provided in \secref{sec:converseproofs}. Finally, we conclude the paper in \secref{sec:conclusions}. The encoder error analysis is relegated to \appref{appendix:errorencodeC1}.

\section{System model and notation}\label{sec:systemmodel}
The channels belonging to $\mathrm{Class~I}$, $\mathrm{Class~II}$ and $\mathrm{Class~III}$ are denoted $\mathrm{C}_1$, $\mathrm{C}_2$ and $\mathrm{C}_3$, respectively. Calligraphic letters are used to denote finite sets, with a probability function defined on them. $\mathrm{N}$ is the number of channel uses, and $\n=1,\dots,\mathrm{N}$ denotes the channel index. Uppercase letters denote random variables (RV), while boldface uppercase letters denote a sequence of RVs. The following notation for a sequence of RVs is useful: $\genericRVS{Y}_1^\mathrm{N}\triangleq (\genericRV{Y}_{1,1}, \dots, \genericRV{Y}_{1,N})$; $\genericRVS{Y}_{1}^{\n-1} \triangleq (\genericRV{Y}_{1,1}, \dots, \genericRV{Y}_{1,\n-1})$; and $\genericRVS{Y}^{\mathrm{N}}_{1,\n+1} \triangleq (\genericRV{Y}_{1,\n+1}, \dots, \genericRV{Y}_{1,\mathrm{N}})$. Lowercase letters are used to denote particular realizations of RVs, and boldface lowercase letters denote vectors. The sender is denoted $\mathrm{S}$ and the receivers are denoted $\mathrm{D}_t$, where $t=1,2$ is the receiver index. Discrete RV $\genericRV{X} \in \mathcal{X}$ and $\genericRV{Y}_t \in \mathcal{Y}_t$ denote the channel input and outputs, respectively. The encoder of $\mathrm{S}$ is supplied with side-information $\genericRVS{W} \in \mathcal{W}^\mathrm{N}$, in a noncausal manner. The channel is assumed to be memoryless and is characterized by the conditional distribution $p(\genericRVS{Y}_1,\genericRVS{Y}_2|\genericRVS{X},\genericRVS{W}) = \prod_{\n=1}^{\mathrm{N}}p(\genericRV{Y}_{1,\n},\genericRV{Y}_{2,\n}|\genericRV{X}_{\n},\genericRV{W}_{\n})$. For sake of brevity, in the remainder of this paper, we use $p(\genericRV{x})$ to denote $p(\genericRV{X}=\genericRV{x})$. Unless otherwise stated, $p(\genericRVS{x})=\prod_{\n=1}^{\mathrm{N}}p(\genericRV{x}_\n)$.

To transmit its messages, $\mathrm{S}$ generates two RVs $\genericRV{M}_{t} \in \mathcal{M}_{t}$, where $\mathcal{M}_{t} = \{1,\dots,2^{\mathrm{N}R_{t}}\}$ denotes a set of message indices. Without loss of generality, $2^{\mathrm{N}R_{t}}$ is assumed to be an integer, with $R_{t}$ being the transmission rate intended to $\mathrm{D}_t$. $\genericRV{M}_{t}$ denotes the message $\mathrm{S}$ intends to transmit to $\mathrm{D}_t$, and is assumed to be independently generated and uniformly distributed over the finite set $\mathcal{M}_{t}$. Integer $\genericRV{m}_{t}\in \mathcal{M}_{t}$ is a particular realization of $\genericRV{M}_{t}$ and denotes the message-index.

Given the conditional distribution characterizing the channel, a $((2^{\mathrm{N}R_{1}},2^{\mathrm{N}R_{2}}),\mathrm{N},P_e^{(\mathrm{N})})$ code for the channels $\mathrm{C}_1$ and $\mathrm{C}_2$ comprises $\mathrm{N}$ encoding functions $f$, such that $\genericRVS{X} = \genericRVS{f}(\genericRV{m}_1,\genericRV{m}_2,\genericRVS{W})$; for the channel $\mathrm{C}_3$, it comprises a stochastic encoder, which is defined by the matrix of conditional probabilities $\phi(\genericRVS{X}|\genericRV{m}_{1},\genericRV{m}_{2},\genericRVS{W})$, such that $\sum_{\genericRVS{X}}\phi(\genericRVS{X}|\genericRV{m}_{1},\genericRV{m}_{2},\genericRVS{W}) = 1$. Here, $\phi(\genericRVS{X}|\genericRV{m}_{1},\genericRV{m}_{2},\genericRVS{W})$ denotes the probability that a pair of message-indices $(\genericRV{m}_{1},\genericRV{m}_{2})$ is encoded as $\genericRVS{X} \in \mathcal{X}^\mathrm{N}$ to be transmitted by $\mathrm{S}$, in the presence of noncausal side-information $\genericRVS{W}$. For all channel models, there are two decoders $g_t: \mathcal{Y}^\mathrm{N}_t \rightarrow \mathcal{M}_{t}$.

The average probability of decoding error for the code, averaged over all codes, is
$P_e^{(\mathrm{N})} = \max \{P_{e,1}^{(\mathrm{N})},P_{e,2}^{(\mathrm{N})}\}$, where, $P_{e,t}^{(\mathrm{N})} = \sum_{\genericRV{m}_{1},\genericRV{m}_{2}}\sum_{\genericRVS{W}\in \mathcal{W}^\mathrm{N}}\frac{1}{2^{\mathrm{N}[R_{1}+R_{2}]}}\text{Pr}\left[g_t(\mathcal{Y}_t^\mathrm{N}) \neq \genericRV{m}_{t}|\genericRV{m}_{1},\genericRV{m}_{2},\genericRVS{W}\right]$. A rate pair $(R_{1},R_{2})$ is said to be achievable for the channel $\mathrm{C}_{\mathrm{c}}; \mathrm{c}=1,2,3$, if there exists a sequence of $((2^{\mathrm{N}R_{1}},2^{\mathrm{N}R_{2}}),\mathrm{N},P_e^{(\mathrm{N})})$ codes, such that $\forall \epsilon > 0$ and sufficiently small,  $P_e^{(\mathrm{N})} \leq \epsilon$ as $\mathrm{N} \rightarrow \infty$. Furthermore, for the channel $\mathrm{C}_3$, the following constraints \cite{Maurer2000} on the conditional entropy must be satisfied for $(R_1,R_2)$ to be considered achievable:
\begin{eqnarray}
\mathrm{N}R_{1} - H(\genericRV{M}_{1}|\genericRVS{Y}_2) \leq \mathrm{N}\epsilon, \label{eq:security1}   \\
\mathrm{N}R_{2} - H(\genericRV{M}_{2}|\genericRVS{Y}_1) \leq \mathrm{N}\epsilon. \label{eq:security2}
\end{eqnarray}
The capacity region is defined as the closure of the set of all achievable rate pairs $(R_{1},R_{2})$.

\section{Main results}\label{sec:statementofresults}
In this section, we state the achievability and converse theorems for all the channel models considered in this paper, and provide related discussion. Let $\mathcal{C}_{\mathrm{c}}$ denote the capacity region of the channel $\mathrm{C}_{\mathrm{c}}$; $\mathrm{c}=1,2,3$. We use the following auxiliary RVs defined on finite sets: $\genericRV{U} \in \mathcal{U}$, $\genericRV{V}_1 \in \mathcal{V}_1$ and $\genericRV{V}_2 \in \mathcal{V}_2$.

\subsection{$\mathrm{Class~I}$ channels}\label{subsec:resultsC1}
For the channel $\mathrm{C}_1$, we consider the set $\mathcal{P}_1$ of all joint probability distributions $p_1(.)$ that can be factored as $p(\genericRV{w})p(\genericRV{v}_1,\genericRV{v}_2|\genericRV{w})p(\genericRV{x}|\genericRV{w},\genericRV{v}_1,\genericRV{v}_2)
p(\genericRV{y}_1,\genericRV{y}_2|\genericRV{x})$. For a given $p_1(.)\in \mathcal{P}_1$, a lower bound on the capacity region for $\mathrm{C}_1$ is described by the set $\mathcal{R}_{1,\text{in}}(p_1)$, which is defined as the union over all distributions $p_1(.)$ of the convex hull of the set of all rate pairs $(R_{1},R_{2})$ that simultaneously satisfy $(\ref{eq:rateregionC1R1})$ - $(\ref{eq:rateregionC1R1plusR2})$.
\begin{eqnarray}
R_1 &\leq& I(\genericRV{V}_1;\genericRV{Y}_1) - I(\genericRV{V}_1;\genericRV{W}),\label{eq:rateregionC1R1}\\
R_2 &\leq& I(\genericRV{V}_2;\genericRV{Y}_2) - I(\genericRV{V}_2;\genericRV{W}),\label{eq:rateregionC1R2}\\
R_1 + R_2 &\leq& I(\genericRV{V}_1;\genericRV{Y}_1) + I(\genericRV{V}_2;\genericRV{Y}_2) - I(\genericRV{V}_1;\genericRV{V}_2) - I(\genericRV{V}_1,\genericRV{V}_2;\genericRV{W}), \label{eq:rateregionC1R1plusR2}
\end{eqnarray}
where $\genericRV{V}_1$ and $\genericRV{V}_2$ are constrained to satisfy the Markov chain $(\genericRV{V}_1,\genericRV{V}_2) \rightarrow (\genericRV{X},\genericRV{W}) \rightarrow (\genericRV{Y}_1,\genericRV{Y}_2)$.
\begin{theorem}\label{thm:achievethmC1}
Let $\mathcal{R}_{1,\text{in}} = \bigcup_{p_1(.)\in \mathcal{P}_1}\mathcal{R}_{1,\text{in}}(p_1)$. Then, $\mathcal{R}_{1,\text{in}}\subseteq \mathcal{C}_1$.
\end{theorem}
For proof, see \secref{subsec:achieveproofC1}.

For a given $p_1(.)\in \mathcal{P}_1$, an outer bound for $\mathrm{C}_1$ is described by the set $\mathcal{R}_{1,\text{out}}(p_1)$, which is defined as the union of all rate pairs $(R_{1},R_{2})$ that simultaneously satisfy $(\ref{eq:outboundC1R1})$ - $(\ref{eq:outboundC1R2})$.
\begin{eqnarray}
R_1 &\leq& I(\genericRV{V}_1;\genericRV{Y}_1) - I(\genericRV{V}_1;\genericRV{W}), \label{eq:outboundC1R1}\\
R_2 &\leq& I(\genericRV{V}_2;\genericRV{Y}_2) - I(\genericRV{V}_2;\genericRV{W}), \label{eq:outboundC1R2},
\end{eqnarray}
where $(\genericRV{V}_1,\genericRV{V}_2) \rightarrow (\genericRV{X},\genericRV{W}) \rightarrow (\genericRV{Y}_1,\genericRV{Y}_2)$.
\begin{theorem}\label{thm:conversethmC1}
Let $\mathcal{R}_{1,\text{out}} = \bigcup_{p_1(.)\in \mathcal{P}_1}\mathcal{R}_{1,\text{out}}(p_1)$. Then, $ \mathcal{C}_{1} \subseteq \mathcal{R}_{1,\text{out}}$.
\end{theorem}
The proof of \thrmref{thm:conversethmC1} can be found in \secref{subsec:converseproofC1}. However, this outer bound does not include a bound on the sum-rates. To explicitly bound the sum-rate, we provide the following alternative outer bound for the channel $\mathrm{C}_1$. We consider the set $\mathcal{P}^{\ast}_1$ of all joint probability distributions $p^{\ast}_1(.)$ that can be factorized as follows: $p(w)p(u,v_1,v_2|w)p(x|w,u,v_1,v_2)p(y_1,y_2|x)$. For a given $p^{\ast}_1(.)\in \mathcal{P}^{\ast}_1$, an outer bound for $\mathrm{C}_1$ is described by the set $\mathcal{R}^{\ast}_{1,\text{out}}(p^{\ast}_1)$, which is defined as the union of all rate pairs $(R_{1},R_{2})$ that simultaneously satisfy \eqref{eq:oboundnewC1R1} - \eqref{eq:oboundnewC1R1plusR2_2}.
\begin{eqnarray}
R_1 &\leq& I(\genericRV{U},\genericRV{V}_1;\genericRV{Y}_1) - I(\genericRV{V}_1;\genericRV{W}|\genericRV{U}), \label{eq:oboundnewC1R1}\\
R_2 &\leq& I(\genericRV{U},\genericRV{V}_2;\genericRV{Y}_2) - I(\genericRV{V}_2;\genericRV{W}|\genericRV{U}), \label{eq:oboundnewC1R2}\\
R_1 + R_2 &\leq& I(\genericRV{U},\genericRV{V}_1;\genericRV{Y}_1) - I(\genericRV{V}_1;\genericRV{W}|\genericRV{U}) + I(\genericRV{U},\genericRV{V}_2;\genericRV{Y}_2|\genericRV{V}_1) - I(\genericRV{V}_2;\genericRV{W}|\genericRV{U},\genericRV{V}_1),\label{eq:oboundnewC1R1plusR2_1}\\
R_1 + R_2 &\leq& I(\genericRV{U},\genericRV{V}_2;\genericRV{Y}_2) - I(\genericRV{V}_2;\genericRV{W}|\genericRV{U}) + I(\genericRV{U},\genericRV{V}_1;\genericRV{Y}_1|\genericRV{V}_2) - I(\genericRV{V}_1;\genericRV{W}|\genericRV{U},\genericRV{V}_2),\label{eq:oboundnewC1R1plusR2_2}
\end{eqnarray}
where the following Markov chain is satisfied: $(\genericRV{U},\genericRV{V}_1,\genericRV{V}_2) \rightarrow (\genericRV{X},\genericRV{W}) \rightarrow (\genericRV{Y}_1,\genericRV{Y}_2)$.
\begin{theorem}\label{thm:conversethmC12}
Let $\mathcal{R}^{\ast}_{1,\text{out}} = \bigcup_{p^{\ast}_1(.)\in \mathcal{P}^{\ast}_1}\mathcal{R}^{\ast}_{1,\text{out}}(p^{\ast}_1)$. Then, $ \mathcal{C}_{1} \subseteq \mathcal{R}^{\ast}_{1,\text{out}}$.
\end{theorem}
\secref{subsec:converseproofC12} contains the proof of \thrmref{thm:conversethmC12}.

\subsection{$\mathrm{Class~II}$ channels}\label{subsec:resultsC2}
For the channel $\mathrm{C}_2$, we consider the set $\mathcal{P}_2$ of all joint probability distributions $p_2(.)$ of the form $p(\genericRV{w})p(\genericRV{u}|\genericRV{w})p(\genericRV{x}|\genericRV{w},\genericRV{u})p(\genericRV{y}_1,\genericRV{y}_2|\genericRV{x})$. For a given $p_2(.)\in \mathcal{P}_2$, a lower bound on the capacity region for $\mathrm{C}_2$ is described by the set $\mathcal{R}_{2,\text{in}}(p_2)$, which is defined as the union over all distributions $p_2(.)$ of the convex-hull of the set of all rate pairs $(R_{1},R_{2})$ that simultaneously satisfy $(\ref{eq:rateregionC2R1})$ - $(\ref{eq:rateregionC2R2})$.
\begin{eqnarray}
R_1 &\leq& I(\genericRV{U};\genericRV{Y}_1) - I(\genericRV{U};\genericRV{W}),\label{eq:rateregionC2R1}\\
R_2 &\leq& I(\genericRV{U};\genericRV{Y}_2) - I(\genericRV{U};\genericRV{W}),\label{eq:rateregionC2R2}
\end{eqnarray}
where the Markov chain $\genericRV{U} \rightarrow (\genericRV{X},\genericRV{W}) \rightarrow (\genericRV{Y}_1,\genericRV{Y}_2)$ holds.
\begin{theorem}\label{thm:achievethmC2}
Let $\mathcal{R}_{2,\text{in}} = \bigcup_{p_2(.)\in \mathcal{P}_2}\mathcal{R}_{2,\text{in}}(p_2)$. Then, $\mathcal{R}_{2,\text{in}}\subseteq \mathcal{C}_2$.
\end{theorem}
The proof of \thrmref{thm:achievethmC2} is relegated to \secref{subsec:achieveproofC2}.

For a given $p_2(.)\in \mathcal{P}_2$, an outer bound for $\mathrm{C}_2$ is described by the set $\mathcal{R}_{2,\text{out}}(p_2)$, which is defined as the union of all rate pairs $(R_{1},R_{2})$ that simultaneously satisfy $(\ref{eq:outboundC2R1})$ - $(\ref{eq:outboundC2R2})$.
\begin{eqnarray}
R_1 &\leq& I(\genericRV{U};\genericRV{Y}_1) - I(\genericRV{U};\genericRV{W}) + H(\genericRV{U}), \label{eq:outboundC2R1}\\
R_2 &\leq& I(\genericRV{U};\genericRV{Y}_2) - I(\genericRV{U};\genericRV{W}) + H(\genericRV{U}),  \label{eq:outboundC2R2}
\end{eqnarray}
with $\genericRV{U} \rightarrow (\genericRV{X},\genericRV{W}) \rightarrow (\genericRV{Y}_1,\genericRV{Y}_2)$.
\begin{theorem}\label{thm:conversethmC2}
Let $\mathcal{R}_{2,\text{out}} = \bigcup_{p_2(.)\in \mathcal{P}_2}\mathcal{R}_{2,\text{out}}(p_2)$. Then, $ \mathcal{C}_{2} \subseteq \mathcal{R}_{2,\text{out}}$.
\end{theorem}
The proof of \thrmref{thm:conversethmC2} can be found in \secref{subsec:converseproofC2}.

\subsection{$\mathrm{Class~III}$ channels}
For the channel $\mathrm{C}_3$, we consider the set $\mathcal{P}_3$ of all joint probability distributions $p_3(.)$ that can be written as $p(\genericRV{w})p(\genericRV{u})p(\genericRV{v}_1,\genericRV{v}_2|\genericRV{w},\genericRV{u})
p(\genericRV{x}|\genericRV{w},\genericRV{v}_1,\genericRV{v}_2)p(\genericRV{y}_1,\genericRV{y}_2|\genericRV{x})$. For a given $p_3(.)\in \mathcal{P}_3$, an inner bound on the capacity region for $\mathrm{C}_3$ is described by the set $\mathcal{R}_{3,\text{in}}(p_3)$, which is defined as the union over all distributions $p_3(.)$ of the convex-hull of the set of all rate pairs $(R_{1},R_{2})$ that simultaneously satisfy $(\ref{eq:rateregionC3R1})$ - $(\ref{eq:rateregionC3R1plusR2})$.
\begin{eqnarray}
R_1 &\leq& I(\genericRV{V}_1;\genericRV{Y}_1|\genericRV{U}) - \max [I(\genericRV{V}_1;\genericRV{Y}_2|\genericRV{U},\genericRV{V}_2),I(\genericRV{V}_1;\genericRV{W}|
\genericRV{U})],\label{eq:rateregionC3R1}\\
R_2 &\leq& I(\genericRV{V}_2;\genericRV{Y}_2|\genericRV{U}) - \max [ I(\genericRV{V}_2;\genericRV{Y}_1|\genericRV{U},\genericRV{V}_1),I(\genericRV{V}_2;\genericRV{W}|
\genericRV{U})],\label{eq:rateregionC3R2}\\
\nonumber R_1 + R_2 &\leq& I(\genericRV{V}_1;\genericRV{Y}_1|\genericRV{U}) + I(\genericRV{V}_2;\genericRV{Y}_2|\genericRV{U}) - I(\genericRV{V}_1;\genericRV{Y}_2|\genericRV{U},\genericRV{V}_2) - I(\genericRV{V}_2;\genericRV{Y}_1|\genericRV{U},\genericRV{V}_1)\\
 &&- I(\genericRV{V}_1;\genericRV{V}_2|\genericRV{U}) - I(\genericRV{V}_1,\genericRV{V}_2;\genericRV{W}|\genericRV{U}), \label{eq:rateregionC3R1plusR2}
\end{eqnarray}
where the following Markov chain is satisfied: $\genericRV{U}\rightarrow (\genericRV{V}_1,\genericRV{V}_2) \rightarrow (\genericRV{X},\genericRV{W}) \rightarrow (\genericRV{Y}_1,\genericRV{Y}_2)$.
\begin{theorem}\label{thm:achievethmC3}
Let $\mathcal{R}_{3,\text{in}} = \bigcup_{p_3(.)\in \mathcal{P}_3}\mathcal{R}_{3,\text{in}}(p_3)$. Then, $\mathcal{R}_{3,\text{in}}\subseteq \mathcal{C}_3$.
\end{theorem}
\secref{subsec:achieveproofC3} contains the proof of \thrmref{thm:achievethmC3}.

For a given $p_3(.)\in \mathcal{P}_3$, an outer bound for $\mathrm{C}_3$ is described by the set $\mathcal{R}_{3,\text{out}}(p_3)$, which is defined as the union of all rate pairs $(R_{1},R_{2})$ that simultaneously satisfy $(\ref{eq:minoutboundC3R1})$ - $(\ref{eq:minoutboundC3R2})$.
\begin{eqnarray}
R_1 &\leq& \min[I_1, I^{\ast}_1], \label{eq:minoutboundC3R1}\\
R_2 &\leq& \min[I_2, I^{\ast}_2], \label{eq:minoutboundC3R2},
\end{eqnarray}
where $I_1,\dots,I^{\ast}_{2}$ are given by $(\ref{eq:outboundC3R1})$ - $(\ref{eq:outboundC3R2genie})$, respectively. \begin{eqnarray}
I_1 &\triangleq& I(\genericRV{V}_1;\genericRV{Y}_1|\genericRV{U}) - I(\genericRV{V}_1;\genericRV{Y}_2|\genericRV{U}) + H(\genericRV{W}|\genericRV{U},V_1), \label{eq:outboundC3R1}\\
I_2 &\triangleq& I(\genericRV{V}_2;\genericRV{Y}_2|\genericRV{U}) - I(\genericRV{V}_2;\genericRV{Y}_1|\genericRV{U}) + H(\genericRV{W}|\genericRV{U},V_2), \label{eq:outboundC3R2}\\
I^{\ast}_1 &\triangleq& I(\genericRV{V}_1;\genericRV{Y}_1|\genericRV{U},\genericRV{V}_2) - I(\genericRV{V}_1;\genericRV{Y}_2|\genericRV{U},\genericRV{V}_2) + H(\genericRV{W}|\genericRV{U},\genericRV{V}_1,\genericRV{V}_2), \label{eq:outboundC3R1genie}\\
I^{\ast}_2 &\triangleq& I(\genericRV{V}_2;\genericRV{Y}_2|\genericRV{U},\genericRV{V}_1) - I(\genericRV{V}_2;\genericRV{Y}_1|\genericRV{U},\genericRV{V}_1) + H(\genericRV{W}|\genericRV{U},\genericRV{V}_1,\genericRV{V}_2), \label{eq:outboundC3R2genie}
\end{eqnarray}
where $\genericRV{U}\rightarrow (\genericRV{V}_1,\genericRV{V}_2) \rightarrow (\genericRV{X},\genericRV{W}) \rightarrow (\genericRV{Y}_1,\genericRV{Y}_2)$. The expressions $(\ref{eq:outboundC3R1genie})$ - $(\ref{eq:outboundC3R2genie})$ are obtained by letting a genie give $\mathrm{D}_1$ message $\genericRV{M}_2$, while $\mathrm{D}_2$ computes the equivocation using $\genericRV{M}_2$ as side-information.
\begin{theorem}\label{thm:conversethmC31}
Let $\mathcal{R}_{3,\text{out}} = \bigcup_{p_3(.)\in \mathcal{P}_3}\mathcal{R}_{3,\text{out}}(p_3)$. Then, $ \mathcal{C}_{3} \subseteq \mathcal{R}_{3,\text{out}}$.
\end{theorem}
The proof of \thrmref{thm:conversethmC31} can be found in \secref{subsec:converseproofC31}. We also provide the following outer bound for the channel $\mathrm{C}_3$, which explicitly characterizes the sum-rates. Consider the set $\mathcal{P}^{\ast}_3$ of all joint probability distributions $p^{\ast}_3(.)$ that can be factorized as follows: $p(w)p(u,v_1,v_2|w)p(x|w,u,v_1,v_2)\\p(y_1,y_2|x)$. For a given $p^{\ast}_3(.)\in \mathcal{P}^{\ast}_3$, an outer bound for $\mathrm{C}_3$ is described by the set $\mathcal{R}^{\ast}_{3,\text{out}}(p^{\ast}_3)$, which is defined as the union of all rate pairs $(R_{1},R_{2})$ that simultaneously satisfy \eqref{eq:oboundnewC3R1} - \eqref{eq:oboundnewC3R1plusR2_2}.
\begin{eqnarray}
R_1 &\leq& I(\genericRV{U},\genericRV{V}_1;\genericRV{Y}_1) - I(\genericRV{V}_1;\genericRV{W}|\genericRV{U}) - I(\genericRV{V}_1;\genericRV{Y}_2),\label{eq:oboundnewC3R1} \\
R_2 &\leq& I(\genericRV{U},\genericRV{V}_2;\genericRV{Y}_2) - I(\genericRV{V}_2;\genericRV{W}|\genericRV{U}) - I(\genericRV{V}_2;\genericRV{Y}_1),\label{eq:oboundnewC3R2} \\
\nonumber R_1 + R_2 &\leq& I(\genericRV{U},\genericRV{V}_1;\genericRV{Y}_1) - I(\genericRV{V}_1;\genericRV{W}|\genericRV{U}) + I(\genericRV{U},\genericRV{V}_2;\genericRV{Y}_2|\genericRV{V}_1) \\ && - I(\genericRV{V}_2;\genericRV{W}|\genericRV{U},\genericRV{V}_1) - I(\genericRV{V}_1;\genericRV{Y}_2),\label{eq:oboundnewC3R1plusR2_1}\\
\nonumber R_1 + R_2 &\leq& I(\genericRV{U},\genericRV{V}_2;\genericRV{Y}_2) - I(\genericRV{V}_2;\genericRV{W}|\genericRV{U}) + I(\genericRV{U},\genericRV{V}_1;\genericRV{Y}_1|\genericRV{V}_2) \\ && - I(\genericRV{V}_1;\genericRV{W}|\genericRV{U},\genericRV{V}_2) - I(\genericRV{V}_2;\genericRV{Y}_1),\label{eq:oboundnewC3R1plusR2_2}
\end{eqnarray}
where $(\genericRV{U},\genericRV{V}_1,\genericRV{V}_2) \rightarrow (\genericRV{X},\genericRV{W}) \rightarrow (\genericRV{Y}_1,\genericRV{Y}_2)$. 
\begin{theorem}\label{thm:conversethmC32}
Let $\mathcal{R}^{\ast}_{3,\text{out}} = \bigcup_{p^{\ast}_3(.)\in \mathcal{P}^{\ast}_3}\mathcal{R}^{\ast}_{3,\text{out}}(p^{\ast}_3)$. Then, $ \mathcal{C}_{3} \subseteq \mathcal{R}^{\ast}_{3,\text{out}}$.
\end{theorem}
The proof of \thrmref{thm:conversethmC32} can be found in \secref{subsec:converseproofC32}.


\subsection{Discussion}\label{subsec:discussion}
\begin{figure}[t]
\centering
\includegraphics[height=2.5in,width=2.75in]{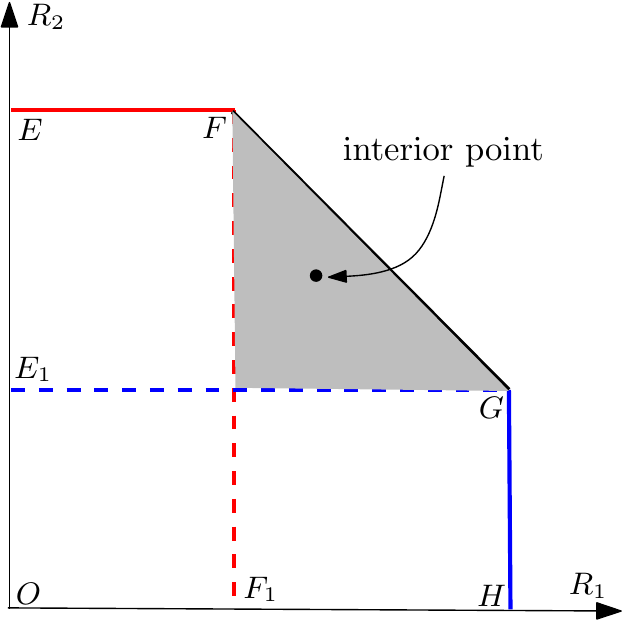}
\caption{Pictorial representation of the rate region for $\mathrm{Class~I}$ channels.}\label{fig:C1_bc1}
\end{figure}

A pictorial representation of the rate region for the channel $\mathrm{C}_1$ is shown in \figref{fig:C1_bc1}. When $R_2=0$, the channel resembles a single-user channel $(\mathrm{S},\mathrm{D}_1)$ with side-information (the Gel'fand-Pinsker's (GP) channel \cite{Gel'fand1980}) and $\mathrm{S}$ can transmit at the maximum achievable $R_1$ given by $(\ref{eq:rateregionC1R1})$, denoted by point the $\mathrm{H}$. At the point $\mathrm{H}$, the maximum achievable $R_2$ is given by the point $\mathrm{E}_1 \equiv I(\genericRV{V}_2;\genericRV{Y}_2) - I(\genericRV{V}_1;\genericRV{V}_2) - I(\genericRV{W};\genericRV{V}_2)$; this is obtained by treating the channel $(\mathrm{S},\mathrm{D}_2)$ as a single-user channel with side-information. Therefore, the rectangle $\mathrm{OHGE}_1$ is achievable. By exchanging $R_1$ and $R_2$ and following similar arguments the points $\mathrm{E}$, given by $(\ref{eq:rateregionC1R2})$, and $\mathrm{F}_1 \equiv I(\genericRV{V}_1;\genericRV{Y}_1) - I(\genericRV{V}_1;\genericRV{V}_2|\genericRV{U}) - I(\genericRV{W};\genericRV{V}_1)$ are achievable. Hence, the rectangle $\mathrm{OEFF}_1$ is also achievable. Since the points $\mathrm{F}$ and $\mathrm{G}$ are shown to be achievable, any point which lies on the line $\mathrm{FG}$ can also be achieved by deriving a bound on the binning rates (see $(\ref{eq:appenderrorencodeC1R1'})$ - $(\ref{eq:appenderrorencodeC1R1plusR2'})$, \appref{appendix:errorencodeC1}). This leads to a sum rate bound given by $(\ref{eq:rateregionC1R1plusR2})$. Finally, owing to convexity of the rate region, any point in the interior of the line $\mathrm{FG}$ is also achievable. Therefore, an achievable rate region for $\mathrm{C}_1$ is described by the pentagon $\mathrm{OEFGH}$.

In the absence of side-information, \ie, $\mathcal{W}=\{\phi\}$, the channel reduces to the classical two-user BC whose rate region is described by the convex-hull of the set of all rate pairs $(R_1,R_2)$ that satisfy the following inequalities:
\begin{eqnarray}
R_1 &\leq& I(\genericRV{V}_1;\genericRV{Y}_1),\label{eq:rateregionBCR1}\\
R_2 &\leq& I(\genericRV{V}_2;\genericRV{Y}_2),\label{eq:rateregionBCR2}\\
R_1 + R_2 &\leq& I(\genericRV{V}_1;\genericRV{Y}_1) + I(\genericRV{V}_2;\genericRV{Y}_2) - I(\genericRV{V}_1;\genericRV{V}_2). \label{eq:rateregionBCR1plusR2}
\end{eqnarray}

For channels of $\mathrm{Class~II}$, each bound in $\eqref{eq:rateregionC2R1}$ - $\eqref{eq:rateregionC2R2}$ is the capacity of GP's single-user channel with noncausal side-information. In the absence of side-information, \ie, $\mathcal{W}=\{\phi\}$, we get $R_t \leq I(U;Y_t) = I(X;Y_t)$, which represents the capacity region of BC when each receiver is given the message it need not decode \cite{Kramer2007}. Furthermore, the outer bounds $\eqref{eq:outboundC2R1}$ - $\eqref{eq:outboundC2R2}$ is within a fixed gap, $H(\genericRV{U})$, from the achievable region, where $H(\genericRV{U})$ is independent of the distribution characterizing this class of channels.

For $\mathrm{Class~III}$ channels, the terms $I(\genericRV{V}_1;\genericRV{Y}_2|\genericRV{U},\genericRV{V}_2)$ and $I(\genericRV{V}_2;\genericRV{Y}_1|\genericRV{U},\genericRV{V}_1)$ quantify the rate-penalty for having to deal with confidentiality constraints on the messages, while the terms $I(\genericRV{V}_1;\genericRV{W}|\genericRV{U})$ and $I(\genericRV{V}_2;\genericRV{W}|\genericRV{U})$ quantify the rate-penalty for having to deal with side-information.

\begin{figure}[t]
\centering
\includegraphics[height=2.5in,width=2.75in]{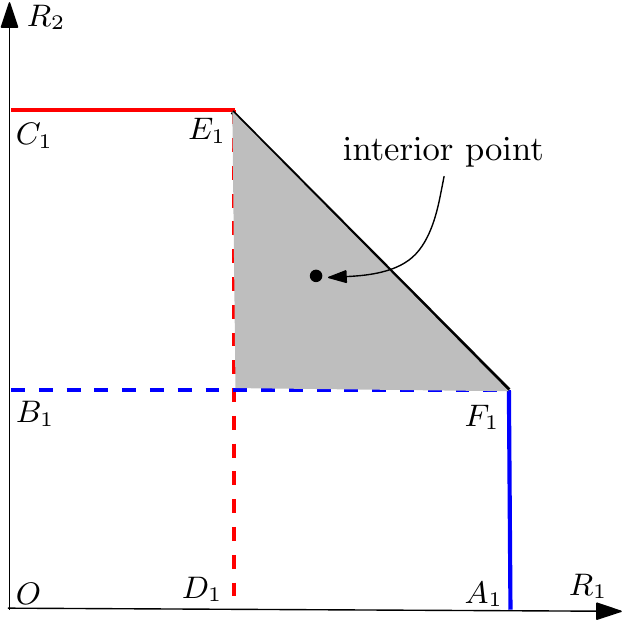}
\caption{Pictorial representation of the rate region for $\mathrm{Class~III}$ channels.}\label{fig:C1C3_bc1}
\end{figure}

Using a combination of results from GP's channel and wiretap channels with side-information \cite{Chen2008}, we obtain a pictorial representation of the rate region for the channel $\mathrm{C}_3$  as shown in \figref{fig:C1C3_bc1}. The arguments used to obtain this schematic are similar to those used for the channel $\mathrm{C}_1$; therefore, we briefly explain the construction of \figref{fig:C1C3_bc1}. The point $\mathrm{A}_1$ corresponds to the maximum achievable $R_1$ (when $R_2=0$) and is given by $(\ref{eq:rateregionC3R1})$. Exchanging $R_1$ and $R_2$ we get the point $\mathrm{C}_1$ given by $(\ref{eq:rateregionC3R2})$. The points $\mathrm{B}_1 \equiv I(\genericRV{V}_2;\genericRV{Y}_2|\genericRV{U}) - I(\genericRV{V}_2;\genericRV{Y}_1|\genericRV{U},\genericRV{V}_1) - \max [I(\genericRV{V}_1;\genericRV{V}_2|\genericRV{U}) ,I(\genericRV{W};\genericRV{V}_2|\genericRV{U})]$ and $\mathrm{D}_1 \equiv I(\genericRV{V}_1;\genericRV{Y}_1|\genericRV{U}) - I(\genericRV{V}_1;\genericRV{Y}_2|\genericRV{U},\genericRV{V}_2) - \max [I(\genericRV{V}_1;\genericRV{V}_2|\genericRV{U}) ,I(\genericRV{W};\genericRV{V}_1|\genericRV{U})]$ are achievable by treating channels $(\mathrm{S},\mathrm{D}_2)$ and $(\mathrm{S},\mathrm{D}_1)$, respectively, as wiretap channels with side-information. The line $\mathrm{E}_1\mathrm{F}_1$ corresponds to the sum rate bound given by $(\ref{eq:rateregionC3R1plusR2})$. Finally, owing to convexity of the rate region, any point in the interior of the line $\mathrm{E}_1\mathrm{F}_1$ is also achievable. Therefore, an achievable rate region for $\mathrm{C}_3$ is described by the pentagon $\mathrm{O}\mathrm{A}_1\mathrm{F}_1\mathrm{E}_1\mathrm{C}_1$.

If the confidentiality constraints $(\ref{eq:security1})$ - $(\ref{eq:security2})$ are relaxed, the channel $\mathrm{C}_3$ reduces to the channel $\mathrm{C}_1$, whose rate region is described by $(\ref{eq:rateregionC1R1})$ - $(\ref{eq:rateregionC1R1plusR2})$. Further, in the absence of side-information, \ie, $\mathcal{W}=\{\phi\}$, the channel reduces to the classical two-user BC whose rate region is described by $\eqref{eq:rateregionBCR1}$ -  $\eqref{eq:rateregionBCR1plusR2}$. Lastly, if the encoder satisfies confidentiality constraints in the absence of side-information, the channel $\mathrm{C}_3$ reduces to BC with two independent and confidential messages whose rate region was first characterized by Liu {\etal} \cite{Liu2008}. It is described by the convex-hull of the set of all rate pairs $(R_1,R_2)$ that satisfy the following inequalities:
\begin{eqnarray}
R_1 \leq I(\genericRV{V}_1;\genericRV{Y}_1|\genericRV{U}) - I(\genericRV{V}_1;\genericRV{Y}_2|\genericRV{U}) - I(\genericRV{V}_1;\genericRV{V}_2|\genericRV{U}),\label{eq:rateregionBCconfR1}\\
R_2 \leq I(\genericRV{V}_2;\genericRV{Y}_2|\genericRV{U}) - I(\genericRV{V}_2;\genericRV{Y}_1|\genericRV{U}) - I(\genericRV{V}_1;\genericRV{V}_2|\genericRV{U}).\label{eq:rateregionBCconfR2}
\end{eqnarray}

\subsection{Relation to past work}\label{subsec:pastwork}
For $\mathrm{Class~I}$ channels, an inner bound was presented in \cite{Steinberg2005a} by extending Marton's achievability scheme for the classical two-user BC to include noncausal side-information at the encoder. In this paper, we employ Marton's technique and use results from the second moment method \cite{Alon2000} to derive the inner bound which matches with the results presented in \cite{Steinberg2005a}. However, our method is simpler and generalizes well for obtaining inner bounds with other channel models, {\eg}, for channels of $\mathrm{Class~III}$ considered in this paper. For the outer bound (specifically, for the sum-rate), we generalize the technique presented in \cite{Nair2007}, to handle side-information at the encoder. When the side-information constraint is relaxed, our result reduces to the one presented for the classical two-user BC \cite{Nair2007}.

$\mathrm{Class~II}$ channels were also addressed in \cite{Oechtering2009}, where an inner bound was derived by employing Marton's achievability scheme. An outer bound was also suggested in \cite{Oechtering2009}, but without a formal proof. In this paper, we derive an inner bound by generalizing the method suggested in \cite{Kramer2007} by incorporating noncausal side-information at the encoder. Our inner bound coincides with the one presented in \cite{Oechtering2009}, but once again the proof technique is much simpler. Furthermore, for the outer bounds, we explicitly address the problem of dealing with the two-dimensional rate region with a single auxiliary random variable.

For $\mathrm{Class~III}$ channels, we show that when the confidentiality constraints are relaxed, our achievable rate region reduces to region presented for the $\mathrm{Class~I}$ channels, and hence to the one presented in \cite{Steinberg2005a}. On the other hand, in the absence of side-information, our achievable region includes an explicit bound on the sum-rate for the two-user BC with confidentiality constraints (a model considered in \cite{Liu2008}). This further strengthens the generalization of our proof technique.


\section{Proofs of achievability theorems}\label{sec:achieveproof}
In this section, we prove \thrmref{thm:achievethmC1}, \thrmref{thm:achievethmC2} and \thrmref{thm:achievethmC3}. For any $\epsilon > 0$, we denote by $A^{(\mathrm{N})}_{\epsilon}(P_{\genericRV{X}})$ an $\epsilon$-typical set comprising sequences picked from the distribution $p(\genericRVS{x})$.  For all the channel models, the encoder is given an $\epsilon-$typical sequence $\genericRVS{W} \in A^{(\mathrm{N})}_{\epsilon}(P_{\genericRV{W}})$ in a noncausal manner.

\subsection{Proof of \thrmref{thm:achievethmC1}}\label{subsec:achieveproofC1}
For the channel $\mathrm{C}_1$, generate $2^{\mathrm{N}[R_t+R^{'}_t]}$ independent typical sequences $\genericRVS{V}_t(i_t,j_t) \in A^{(\mathrm{N})}_{\epsilon}(P_{\genericRV{V}_t}); t = 1,2$. Here, $i_t\in\{1,\dots,2^{\mathrm{N}R_t}\}$; $j_t\in\{1,\dots,2^{\mathrm{N}R^{'}_t}\}$. Uniformly distribute $2^{\mathrm{N}[R_t+R^{'}_t]}$ sequences into $2^{\mathrm{N}R_t}$ bins, so that each bin, indexed by $i_t$, comprises $2^{\mathrm{N}R^{'}_t}$ sequences. To send the message pair $(\genericRV{m}_1 = i_1, \genericRV{m}_2 = i_2)$, the encoder at $\mathrm{S}$ looks for a pair $(j_1,j_2)$ that satisfies the following joint typicality condition: $E_{\mathrm{S}} \triangleq \{(\genericRVS{W},\genericRVS{V}_1(i_1,j_1),\genericRVS{V}_2(i_2,j_2))\in A^{(\mathrm{N})}_{\epsilon}(P_{\genericRV{W},\genericRV{V}_1,\genericRV{V}_2})\}$. An error is declared at the encoder of $\mathrm{S}$, if it is not possible to find the $(j_1,j_2)-$pair to satisfy the condition $E_{\mathrm{S}}$. The encoder error analysis can be found in \appref{appendix:errorencodeC1}. The channel input sequence is $\genericRVS{X} \in A^{(\mathrm{N})}_{\epsilon}(P_{\genericRV{X|W,V_1,V_2}})$.

At the destination $\mathrm{D}_t$, the decoder looks for $(\hat{i}_t,\hat{j}_t)$ that satisfies the following joint typicality condition: $E_{\mathrm{D}_t} \triangleq \{(\genericRVS{V}_t(\hat{i}_t,\hat{j}_t),\genericRVS{Y}_t)\in A^{(\mathrm{N})}_{\epsilon}(P_{\genericRV{V}_t,\genericRV{Y}_t})\}$. An error is declared at decoder of $\mathrm{D}_t$, if it not possible to find a unique integer $\hat{i}_t$ to satisfy the condition $E_{\mathrm{D}_t}$. From the union of events bound, the probability of decoder error at $\mathrm{D}_t$ can be upper bounded as follows: $P^{(\mathrm{N})}_{e,\mathrm{D}_t} \leq \text{Pr}(E^c_{\mathrm{D}_t}|E_{\mathrm{S}}) + \sum_{\hat{i}_t\neq i_t}\sum_{j_t}\text{Pr}(E_{\mathrm{D}_t}|E_{\mathrm{S}})$. From the asymptotic equipartition property (AEP) \cite{Cover2006}, $\forall \epsilon > 0$ and sufficiently small; and for large $\mathrm{N}$, $\text{Pr}(E^c_{\mathrm{D}_t}|E_{\mathrm{S}}) \leq \epsilon$. Further, for $\hat{i}_t\neq i_t$, $\text{Pr}(E_{\mathrm{D}_t}|E_{\mathrm{S}}) \leq 2^{-\mathrm{N}[I(V_t;Y_t)-\epsilon]}$. Therefore, we have $P^{(\mathrm{N})}_{e,\mathrm{D}_t} \leq \epsilon + 2^{\mathrm{N}[R_t+R^{'}_t]}2^{-\mathrm{N}[I(V_t;Y_t)-\epsilon]}$, leading us to conclude that, for any $\epsilon_0 >0$ and sufficiently small; and for large $\mathrm{N}$, $P^{(\mathrm{N})}_{e,\mathrm{D}_t} \leq \epsilon_0$ if
\begin{eqnarray}
R_t+R^{'}_t < I(\genericRV{V}_t;\genericRV{Y}_t). \label{eq:achieveproofC1rates}
\end{eqnarray}

For the channel $\mathrm{C}_1$, the rate inequalities $(\ref{eq:achieveproofC1rates})$ and the bounds on the binning rates $(\ref{eq:appenderrorencodeC1R1'})$ - $(\ref{eq:appenderrorencodeC1R1plusR2'})$ (see \appref{appendix:errorencodeC1}) are combined to obtain an achievable rate region given by $(\ref{eq:rateregionC1R1})$ - $(\ref{eq:rateregionC1R1plusR2})$. This completes the proof of \thrmref{thm:achievethmC1}.

\subsection{Proof of \thrmref{thm:achievethmC2}}\label{subsec:achieveproofC2}
For the channel $\mathrm{C}_2$, we consider the following two cases.
\begin{enumerate}
\item When $R_1 \leq R_2$: Generate $2^{\mathrm{N}(R_2+R^{\ast})}$ typical sequences $\genericRVS{U}(i,j) \in A^{(\mathrm{N})}_{\epsilon}(P_{\genericRV{U}}); i \in \{1,\dots,2^{\mathrm{N}R_2}\}; \\ j \in \{1,\dots,2^{\mathrm{N}R^{\ast}}\}$. Uniformly distribute these sequences into $2^{\mathrm{N}R_2}$ bins, so that each bin comprises $2^{\mathrm{N}R^{\ast}}$ sequences. The bins are indexed by $i$. Define now the following mappings:
    \begin{eqnarray*}
    \genericRV{m}_t \in \{1,\dots,2^{\mathrm{N}R_t}\} &\mapsto& \mathrm{Int}(\genericRV{m}_t) \in \{0,\dots,2^{\mathrm{N}R_2}-1\}; t=1,2,
    \end{eqnarray*}
    where $\mathrm{Int}(\alpha)$ denotes an integer to represent $\alpha$. To transmit the message pair $(\genericRV{m}_1,\genericRV{m}_2)$, compute $\left(\mathrm{Int}(\genericRV{m}_1)+\mathrm{Int}(\genericRV{m}_2) \mod{2^{\mathrm{N}R_2}}\right)$. By construction, the bin index \\ $i \triangleq \mathrm{Int}^{-1}\left(\mathrm{Int}(\genericRV{m}_1)+\mathrm{Int}(\genericRV{m}_2) \mod{2^{\mathrm{N}R_2}}\right)$. Given the sequence $\genericRVS{W}$, the encoder looks for an integer $j$ to satisfy the following joint typicality condition:
    \begin{eqnarray*}
    (\genericRVS{U}(i,j), \genericRVS{W})\in A^{(\mathrm{N})}_{\epsilon}(P_{\genericRV{W},\genericRV{U}}).
    \end{eqnarray*}
    Finally, $\genericRVS{X} \triangleq \genericRVS{f}(\genericRVS{U}(i,j),\genericRVS{W})$ is transmitted in $\mathrm{N}$ channel uses.

    At receiver $\mathrm{D}_1$, given $\genericRV{m}_2$, the decoder looks for the pair $(\hat{i}\triangleq \hat{\genericRV{m}}_1,\hat{j})$ such that the following joint typicality condition is satisfied:
    \begin{eqnarray*}
    E_{\mathrm{D}_1}\triangleq \{(\genericRVS{U}(\mathrm{Int}^{-1}\left(\mathrm{Int}(\hat{\genericRV{m}}_1)+\mathrm{Int}(\genericRV{m}_2) \mod{2^{\mathrm{N}R_2}}\right),j), \genericRVS{Y}_1)\in A^{(\mathrm{N})}_{\epsilon}(P_{\genericRV{U},\genericRV{Y}_1})\}.
    \end{eqnarray*}
    From AEP, it can be shown that $\text{Pr}(E^c_{\mathrm{D}_1}) \leq \delta_1; \forall \delta_1 >0$ and sufficiently small; and for large $\mathrm{N}$, if $R_1 + R^{\ast} \leq I(\genericRV{U};\genericRV{Y}_1)$. Similarly, it can be shown that $\text{Pr}(E^c_{\mathrm{D}_2}) \leq \delta_2; \forall \delta_2 >0$ and sufficiently small; and for large $\mathrm{N}$, if $R_2 + R^{\ast} \leq I(\genericRV{U};\genericRV{Y}_2)$. Additionally, by following a procedure similar to the one presented in \appref{appendix:errorencodeC1}, we bound the binning rate as follows: $R^{\ast} > I(\genericRV{U};\genericRV{W})$. Therefore, $\genericRV{m}_1$ (resp. $\genericRV{m}_2$) can be reliably decoded at $\mathrm{D}_1$ (resp. $\mathrm{D}_2$) if
    \begin{eqnarray}
    R_1 \leq I(\genericRV{U};\genericRV{Y}_1) - I(\genericRV{U};\genericRV{W}), \label{eq:achieveproofC2R1}\\
    R_2 \leq I(\genericRV{U};\genericRV{Y}_2) - I(\genericRV{U};\genericRV{W}). \label{eq:achieveproofC2R2}
    \end{eqnarray}

\item When $R_2 \leq R_1$: By symmetry, we get the same rate bounds as in $\eqref{eq:achieveproofC2R1}$ and $\eqref{eq:achieveproofC2R2}$.
\end{enumerate}
This completes the proof of \thrmref{thm:achievethmC2}.

\subsection{Proof of \thrmref{thm:achievethmC3}}\label{subsec:achieveproofC3}
For the channel $\mathrm{C}_3$, generate a typical sequence $\genericRVS{U} \in A^{(\mathrm{N})}_{\epsilon}(P_{\genericRV{U}})$, known to all nodes in the network. Generate $2^{\mathrm{N}[R_t+R^{'}_t+R_t^{\ast}]}$ independent typical sequences $\genericRVS{V}_t(i_t,j_t,k_t) \in A^{(\mathrm{N})}_{\epsilon}(P_{\genericRV{V}_t})$; $i_t\in\{1,\dots,2^{\mathrm{N}R_t}\}$; $j_t\in\{1,\dots,2^{\mathrm{N}R^{'}_t}\}$; $k_t\in\{1,\dots,2^{\mathrm{N}R_t^{\ast}}\}$. Uniformly distribute $2^{\mathrm{N}[R_t+R^{'}_t+R_t^{\ast}]}$ sequences into $2^{\mathrm{N}R_t}$ bins, so that each bin, indexed by $i_t$, comprises $2^{\mathrm{N}[R^{'}_t+R_t^{\ast}]}$ sequences. Uniformly distribute $2^{\mathrm{N}[R^{'}_t+R_t^{\ast}]}$ sequences into $2^{\mathrm{N}R^{'}_t}$ sub-bins indexed by $(i_t,j_t)$, so that each sub-bin comprises $2^{\mathrm{N}R_t^{\ast}}$ sequences.

To send the message pair $(\genericRV{m}_1,\genericRV{m}_2)$, $\mathrm{S}$ employs a stochastic encoder. In the bin indexed by $i_t$, \emph{randomly} pick a sub-bin indexed $(i_t,j_t)$. The encoder then looks for a pair $(k_1,k_2)$ that satisfies the following joint typicality condition: $(\genericRVS{W},\genericRVS{V}_1(i_1,j_1,k_1),\genericRVS{V}_2(i_2,j_2,k_2))\in A^{(\mathrm{N})}_{\epsilon}(P_{\genericRV{W},\genericRV{V}_1,\genericRV{V}_2|\genericRV{U}})$.
The channel input sequence $\genericRVS{X} \in A^{(\mathrm{N})}_{\epsilon}(P_{\genericRV{X}|\genericRV{W},\genericRV{V}_1,\genericRV{V}_2})$ is transmitted in $\mathrm{N}$ uses of the channel.

At the destination $\mathrm{D}_t$, given $\genericRVS{U}$, the decoder picks $k_t$ that satisfies the following joint typicality condition: $E_{\mathrm{D}_t} \triangleq \{(\genericRVS{V}_t(i_t,j_t,k_t),\genericRVS{Y}_t)\in A^{(\mathrm{N})}_{\epsilon}(P_{\genericRV{V}_t,\genericRV{Y}_t|\genericRV{U}})\}$. An error is declared at the decoder of $\mathrm{D}_t$ if it not possible to find an integer $\hat{i}_t$ satisfying $E_{\mathrm{D}_t}$. From union of events bound, the probability of decoder error at $\mathrm{D}_t$ can be upper bounded as follows: $P^{(\mathrm{N})}_{e,\mathrm{D}_t} \leq \text{Pr}(E^c_{\mathrm{D}_t}|E_{\mathrm{S}}) + \sum_{\hat{i}_t\neq i_t}\sum_{j_t,k_t}\text{Pr}(E_{\mathrm{D}_t}|E_{\mathrm{S}})$. From AEP \cite{Cover2006}, $\forall \epsilon > 0$ and sufficiently small; and for large $\mathrm{N}$, $\text{Pr}(E^c_{\mathrm{D}_t}|E_{\mathrm{S}}) \leq \epsilon$ and for $\hat{i}_t\neq i_t$, we have $\text{Pr}(E_{\mathrm{D}_t}|E_{\mathrm{S}}) \leq 2^{-\mathrm{N}[I(\genericRV{V}_t;\genericRV{Y}_t|U)-\epsilon]}$. Therefore, $P^{(\mathrm{N})}_{e,\mathrm{D}_t} \leq \epsilon + 2^{\mathrm{N}[R_t+R^{'}_t+R_t^{\ast}]}2^{-\mathrm{N}[I(\genericRV{V}_t;\genericRV{Y}_t|\genericRV{U})-\epsilon]}$. For any $\epsilon_0 >0$ and sufficiently small; and for large $\mathrm{N}$, $P^{(\mathrm{N})}_{e,\mathrm{D}_t} \leq \epsilon_0$ if
\begin{eqnarray}
R_t+R^{'}_t+R_t^{\ast} < I(\genericRV{V}_t;\genericRV{Y}_t|\genericRV{U}). \label{eq:achieveproofC3rates}
\end{eqnarray}
The equivocation at the decoder of $\mathrm{D}_2$ is calculated by first considering the following lower bound: $H(M_1|\genericRVS{Y}^{\mathrm{N}}_2) \geq H(M_1|\genericRVS{Y}^{\mathrm{N}}_2,\genericRVS{U}^{\mathrm{N}},\genericRVS{V}^{\mathrm{N}}_2)$. Following the procedure in \cite[Section V-B]{Liu2008} and using the fact that $M_1 \rightarrow (\genericRVS{U}^{\mathrm{N}},\genericRVS{V}^{\mathrm{N}}_1,\genericRVS{V}^{\mathrm{N}}_2) \rightarrow \genericRVS{Y}^{\mathrm{N}}_2$ forms a Markov chain, we get
\begin{eqnarray}
H(M_1|\genericRVS{Y}^{\mathrm{N}}_2)  \geq  H(\genericRVS{V}^{\mathrm{N}}_1|\genericRVS{U}^{\mathrm{N}}) - I(\genericRVS{V}^{\mathrm{N}}_1;\genericRVS{V}^{\mathrm{N}}_2|\genericRVS{U}^{\mathrm{N}})
- H(\genericRVS{V}^{\mathrm{N}}_1|M_1,\genericRVS{U}^{\mathrm{N}},\genericRVS{V}^{\mathrm{N}}_2,\genericRVS{Y}^{\mathrm{N}}_2) -
I(\genericRVS{V}^{\mathrm{N}}_1;\genericRVS{Y}^{\mathrm{N}}_2|\genericRVS{U}^{\mathrm{N}},\genericRVS{V}^{\mathrm{N}}_2). \label{eq:achieveproofC3equiv2}
\end{eqnarray}
Now, $\forall \epsilon_l > 0; l=4,\dots,10$ and sufficiently small; and for large $\mathrm{N}$, the terms in $(\ref{eq:achieveproofC3equiv2})$ become
\begin{eqnarray}
\nonumber H(\genericRVS{V}^{\mathrm{N}}_1|\genericRVS{U}^{\mathrm{N}}) \stackrel{(a)}{=} \mathrm{N}[R_1+R'_1+R^{\ast}_1];
I(\genericRVS{V}^{\mathrm{N}}_1;\genericRVS{V}^{\mathrm{N}}_2|\genericRVS{U}^{\mathrm{N}})\stackrel{(b)}{=}\mathrm{N}
I(\genericRV{V}_1;\genericRV{V}_2|\genericRV{U})
+\mathrm{N}\epsilon_4; \\
H(\genericRVS{V}^{\mathrm{N}}_1|M_1,\genericRVS{U}^{\mathrm{N}},\genericRVS{V}^{\mathrm{N}}_2,\genericRVS{Y}^{\mathrm{N}}_2) \stackrel{(c)}{\leq} \mathrm{N}\epsilon_5;
I(\genericRVS{V}^{\mathrm{N}}_1;\genericRVS{Y}^{\mathrm{N}}_2|\genericRVS{U}^{\mathrm{N}},\genericRVS{V}^{\mathrm{N}}_2) \stackrel{(d)}{=}\mathrm{N}I(\genericRV{V}_1;\genericRV{Y}_2|\genericRV{U},\genericRV{V}_2)+\mathrm{N}\epsilon_6.
\label{eq:achieveproofC3equiv3}
\end{eqnarray}
In \eqref{eq:achieveproofC3equiv3}, $(a)$ follows from the codebook construction; $(b)$ and $(d)$ follow from standard techniques (for \eg, see \cite[Lemma 3]{Liu2008}); and $(c)$ is proved in \cite[Lemma 2]{Liu2008}. A similar procedure is followed to calculate the equivocation at the decoder at $\mathrm{D}_1$. Finally, the security constraints $(\ref{eq:security1})$ and $(\ref{eq:security2})$ are satisfied by letting
\begin{eqnarray}
R'_1 &=& I(\genericRV{V}_1;\genericRV{Y}_2|\genericRV{U},\genericRV{V}_2) - \epsilon_7;
R_1^{\ast} = I(\genericRV{V}_1;\genericRV{V}_2|\genericRV{U}) - \epsilon_8;\label{eq:achieveproofC3R1*}\\
R'_2 &=& I(\genericRV{V}_2;\genericRV{Y}_1|\genericRV{W},\genericRV{U},\genericRV{V}_1) - \epsilon_9;
R_2^{\ast} = I(\genericRV{V}_1;\genericRV{V}_2|\genericRV{W},\genericRV{U}) - \epsilon_{10}. \label{eq:achieveproofC3R2*}
\end{eqnarray}
For the channel $\mathrm{C}_3$, rate inequalities $(\ref{eq:achieveproofC3rates})$, constraints $(\ref{eq:achieveproofC3R1*})$ - $(\ref{eq:achieveproofC3R2*})$ and bounds on the binning rates $\eqref{eq:appenderrorencodeC3R1*}$ - $\eqref{eq:appenderrorencodeC3R1plusR2*}$ (see \appref{appendix:errorencodeC1}) are combined to obtain the rate region described by $(\ref{eq:rateregionC3R1})$ - $(\ref{eq:rateregionC3R1plusR2})$. This completes the proof of \thrmref{thm:achievethmC3}.


\section{Proofs of converse theorems}\label{sec:converseproofs}
In this section, we prove \thrmref{thm:conversethmC1}, \thrmref{thm:conversethmC12}, \thrmref{thm:conversethmC2}, \thrmref{thm:conversethmC31} and \thrmref{thm:conversethmC32}.

\subsection{Proof of \thrmref{thm:conversethmC1}}\label{subsec:converseproofC1}
For the channel $\mathrm{C}_1$, $\forall \epsilon >0$ and sufficiently small; and for large $\mathrm{N}$, $R_1$ can be bounded as follows:
\begin{eqnarray}
\nonumber \mathrm{N}R_1 &=& H(\genericRV{M}_1) = I(\genericRV{M}_1;\genericRVS{Y}_1^\mathrm{N})+H(\genericRV{M}_1|\genericRVS{Y}_1^\mathrm{N})\\
\nonumber &\stackrel{(a)}\leq& I(\genericRV{M}_1;\genericRVS{Y}_1^\mathrm{N}) + \mathrm{N}\epsilon
\stackrel{(b)}= \sum^{\mathrm{N}}_{\n=1}[H(\genericRV{Y}_{1,\n}|\genericRVS{Y}^{\n-1}_1) - H(\genericRV{Y}_{1,\n}|\genericRVS{Y}^{\n-1}_1,\genericRV{M}_1)] + \mathrm{N}\epsilon \\
\nonumber &\stackrel{(c)}\leq& \sum^{\mathrm{N}}_{\n=1}[H(\genericRV{Y}_{1,\n}) - H(\genericRV{Y}_{1,\n}|\genericRVS{Y}^{\n-1}_1,\genericRV{M}_1)] + \mathrm{N}\epsilon
= \sum^{\mathrm{N}}_{\n=1}I(\genericRV{M}_1,\genericRVS{Y}^{\n-1}_1;\genericRV{Y}_{1,\n}) + \mathrm{N}\epsilon \\
\nonumber &=& \sum^{\mathrm{N}}_{\n=1}[I(\genericRV{M}_1,\genericRVS{Y}^{\n-1}_1,\genericRVS{W}^{\mathrm{N}}_{\n+1};\genericRV{Y}_{1,\n}) - I(\genericRVS{W}^{\mathrm{N}}_{\n+1};\genericRV{Y}_{1,\n}|\genericRV{M}_1,\genericRVS{Y}^{\n-1}_1)] + \mathrm{N}\epsilon \\
\nonumber &\stackrel{(d)}=& \sum^{\mathrm{N}}_{\n=1}[I(\genericRV{M}_1,\genericRVS{Y}^{\n-1}_1,\genericRVS{W}^{\mathrm{N}}_{\n+1};\genericRV{Y}_{1,\n}) - I(\genericRVS{Y}^{\n-1}_1;W_\n|\genericRV{M}_1,\genericRVS{W}^{\mathrm{N}}_{\n+1})] + \mathrm{N}\epsilon \\
\nonumber &\stackrel{(e)}=& \sum^{\mathrm{N}}_{\n=1}[I(\genericRV{M}_1,\genericRVS{Y}^{\n-1}_1,\genericRVS{W}^{\mathrm{N}}_{\n+1};\genericRV{Y}_{1,\n}) - I(\genericRV{M}_1,\genericRVS{W}^{\mathrm{N}}_{\n+1},\genericRVS{Y}^{\n-1}_1;W_\n)] + \mathrm{N}\epsilon,
\end{eqnarray}
where $(a)$ follows from Fano's inequality \cite{Cover2006}, $(b)$ follows from the chain rule, $(c)$ follows from the fact that conditioning reduces entropy, $(d)$ follows from Csisz\'{a}r's sum identity \cite{Csisz'ar1982} and $(e)$ is due to the fact that $(\genericRV{M}_1,\genericRVS{W}^{\mathrm{N}}_{\n+1})$ is independent of $\genericRV{W}_\n$.
We let $\genericRV{V}_{1,\n}=(\genericRV{M}_1,\genericRVS{W}^{\mathrm{N}}_{\n+1},\genericRVS{Y}^{\n-1}_1)$ and note that this choice satisfies the Markov chain requirement $\genericRV{V}_1 \rightarrow (\genericRV{X},\genericRV{W}) \rightarrow (\genericRV{Y}_1,\genericRV{Y}_2)$, specified in \secref{sec:statementofresults} for the channel $\mathrm{C}_1$. Thus, we get
\begin{eqnarray}
\mathrm{N}R_1 \leq \sum^{\mathrm{N}}_{\n=1}I(\genericRV{V}_{1,\n};\genericRV{Y}_{1,\n}) - I(\genericRV{V}_{1,\n};\genericRV{W}_\n) + \mathrm{N}\epsilon. \label{eq:converseproofC1R1}
\end{eqnarray}
Proceeding in a similar manner and letting $\genericRV{V}_{2,\n}=(\genericRV{M}_2,\genericRVS{W}^{\mathrm{N}}_{\n+1},\genericRVS{Y}^{\n-1}_2)$, we get
\begin{eqnarray}
\mathrm{N}R_2 \leq \sum^{\mathrm{N}}_{\n=1}I(\genericRV{V}_{2,\n};\genericRV{Y}_{2,\n}) - I(\genericRV{V}_{2,\n};\genericRV{W}_\n) + \mathrm{N}\epsilon. \label{eq:converseproofC1R2}
\end{eqnarray}

\subsection{Proof of \thrmref{thm:conversethmC12}}\label{subsec:converseproofC12}
For the channel $\mathrm{C}_1$, $\forall \epsilon >0$ and sufficiently small; and for large $\mathrm{N}$, $R_1$ can be bounded as
\begin{eqnarray}
\nonumber \mathrm{N}R_1 &=& H(\genericRV{M}_1) = I(\genericRV{M}_1;\genericRVS{Y}_1^\mathrm{N})+H(\genericRV{M}_1|\genericRVS{Y}_1^\mathrm{N})\\
\nonumber &\stackrel{(a)}\leq& I(\genericRV{M}_1;\genericRVS{Y}_1^\mathrm{N}) + \mathrm{N}\epsilon,
\end{eqnarray}
where $(a)$ follows from Fano's inequality. Proceeding in a manner similar to the proof of \thrmref{thm:conversethmC1} (see \secref{subsec:converseproofC1}), and letting $\genericRV{U}_{\mathrm{n}}=(\genericRVS{W}^{\mathrm{N}}_{\mathrm{n}+1},\genericRVS{Y}^{\mathrm{n}-1}_1,
\genericRVS{Y}^{\mathrm{N}}_{2,\mathrm{n}+1})$ and $\genericRV{V}_{1,\mathrm{n}} = \genericRV{M}_1$.
\begin{eqnarray}
\mathrm{N}R_1 \leq \sum^{\mathrm{N}}_{\mathrm{n}=1}I(\genericRV{U}_{\mathrm{n}}, \genericRV{V}_{1,\mathrm{n}};\genericRV{Y}_{1,\mathrm{n}}) - I(\genericRV{V}_{1,\mathrm{n}};\genericRV{W}_{\mathrm{n}}|\genericRV{U}_{\mathrm{n}}) + \mathrm{N}\epsilon. \label{eq:outboundnewC1R1}
\end{eqnarray}
Similarly, letting $\genericRV{V}_{2,\mathrm{n}} = \genericRV{M}_2$, $R_2$ can be upper bounded as follows:
\begin{eqnarray}
\mathrm{N}R_2 \leq \sum^{\mathrm{N}}_{\mathrm{n}=1}I(\genericRV{U}_{\mathrm{n}}, \genericRV{V}_{2,\mathrm{n}};\genericRV{Y}_{2,\mathrm{n}}) - I(\genericRV{V}_{2,\mathrm{n}};\genericRV{W}_{\mathrm{n}}|\genericRV{U}_{\mathrm{n}}) + \mathrm{N}\epsilon. \label{eq:outboundnewC1R2}
\end{eqnarray}

We next upper bound $R_1+R_2$ as follows. $\forall \epsilon >0$ and sufficiently small; and for large $\mathrm{N}$, we have
\begin{eqnarray}
\nonumber \mathrm{N}(R_1+R_2) &=& H(\genericRV{M}_1,\genericRV{M}_2) = H(\genericRV{M}_1) + H(\genericRV{M}_2|\genericRV{M}_1)\\
\nonumber  &=& I(\genericRV{M}_1;\genericRVS{Y}_1^\N)+H(\genericRV{M}_1|\genericRVS{Y}_1^\N) +I(\genericRV{M}_2;\genericRVS{Y}_2^\N|\genericRV{M}_1)+H(\genericRV{M}_2|\genericRVS{Y}_2^\N,\genericRV{M}_1)\\
\nonumber &\stackrel{(a)}\leq& \sum^{\N}_{\n=1}I(\genericRV{M}_1;\genericRV{Y}_{1,\n}|\genericRVS{Y}^{\n-1}_1)+ \sum^{\N}_{\n=1}I(\genericRV{M}_2;\genericRV{Y}_{2,\n}|\genericRVS{Y}^{\N}_{2,\n+1},\genericRV{M}_1)+ 2\N\epsilon,
\end{eqnarray}
where $(a)$ follows from Fano's inequality. Consider
\begin{eqnarray}
\nonumber \sum^{\N}_{\n=1}I(\genericRV{M}_1;\genericRVS{Y}_{1,\n}|\genericRVS{Y}^{\n-1}_1) &\leq& \sum^{\N}_{\n=1}I(\genericRV{M}_1,\genericRVS{Y}^{\n-1}_1;\genericRVS{Y}_{1,\n})\\
\nonumber &=& \sum^{\N}_{\n=1}I(\genericRV{M}_1,\genericRVS{Y}^{\n-1}_1,\genericRVS{Y}^{\N}_{2,\n+1};\genericRVS{Y}_{1,\n})
- \sum^{\N}_{\n=1}I(\genericRVS{Y}^{\N}_{2,\n+1};\genericRVS{Y}_{1,\n}|\genericRV{M}_1,\genericRVS{Y}^{\n-1}_1)\\
\nonumber &=& \sum^{\N}_{\n=1}[I(\genericRV{M}_1,\genericRVS{Y}^{\n-1}_1,\genericRVS{Y}^{\N}_{2,\n+1},\genericRVS{W}^{\N}_{\n+1};\genericRV{Y}_{1,\n}) - I(\genericRVS{W}^{\N}_{\n+1};\genericRV{Y}_{1,\n}|\genericRV{M}_1,\genericRVS{Y}^{\n-1}_1,\genericRVS{Y}^{\N}_{2,\n+1})] \\
\nonumber &-& \sum^{\N}_{\n=1}I(\genericRVS{Y}^{\N}_{2,\n+1};\genericRV{Y}_{1,\n}|\genericRV{M}_1,\genericRVS{Y}^{\n-1}_1)\\
\nonumber &\stackrel{(b)}=& \sum^{\N}_{\n=1}[I(\genericRV{M}_1,\genericRVS{Y}^{\n-1}_1,\genericRVS{Y}^{\N}_{2,\n+1},\genericRVS{W}^{\N}_{\n+1};\genericRV{Y}_{1,\n}) - I(\genericRV{M}_1;\genericRV{W}_{\n}|\genericRVS{W}^{\N}_{\n+1},\genericRVS{Y}^{\n-1}_1,\genericRVS{Y}^{\N}_{2,\n+1})] \\
&-& \sum^{\N}_{\n=1}I(\genericRVS{Y}^{\N}_{2,\n+1};\genericRV{Y}_{1,\n}|\genericRV{M}_1,\genericRVS{Y}^{\n-1}_1)\label{eq:sumrateout1}
\end{eqnarray}

Next consider
\begin{eqnarray}
\nonumber \sum^{\N}_{\n=1}I(M_2;\genericRV{Y}_{2,\n}|\genericRVS{Y}^{\N}_{2,\n+1},\genericRV{M}_1) \leq \sum^{\N}_{\n=1}I(\genericRV{M}_2,\genericRVS{Y}^{\n-1}_1;\genericRV{Y}_{2,\n}|\genericRVS{Y}^{\N}_{2,\n+1},\genericRV{M}_1)\\ \nonumber
= \sum^{\N}_{\n=1}I(\genericRVS{Y}^{\n-1}_1;\genericRV{Y}_{2,\n}|\genericRVS{Y}^{\mathrm{N}}_{2,\n+1},\genericRV{M}_1) +
\sum^{\mathrm{N}}_{\n=1}I(\genericRV{M}_2;\genericRVS{Y}_{2,\n}|\genericRVS{Y}^{\n-1}_1,
\genericRVS{Y}^{\mathrm{N}}_{2,\n+1},\genericRV{M}_1)\\ \nonumber
= \sum^{\mathrm{N}}_{\n=1}I(\genericRVS{Y}^{\n-1}_1;\genericRV{Y}_{2,\n}|\genericRVS{Y}^{\mathrm{N}}_{2,\n+1},\genericRV{M}_1) +
\sum^{\mathrm{N}}_{\n=1}I(\genericRV{M}_2,\genericRVS{W}^{\mathrm{N}}_{\n+1};\genericRV{Y}_{2,\n}|\genericRVS{Y}^{\n-1}_1,
\genericRVS{Y}^{\mathrm{N}}_{2,\n+1},\genericRV{M}_1)\\ \nonumber -  \sum^{\mathrm{N}}_{\n=1}I(\genericRVS{W}^{\mathrm{N}}_{\n+1};\genericRV{Y}_{2,\n}|\genericRVS{Y}^{\n-1}_1,
\genericRVS{Y}^{\mathrm{N}}_{2,\n+1},\genericRV{M}_1,\genericRV{M}_2)\\ \nonumber = \sum^{\mathrm{N}}_{\n=1}I(\genericRVS{Y}^{\n-1}_1;\genericRV{Y}_{2,\n}|\genericRVS{Y}^{\mathrm{N}}_{2,\n+1},\genericRV{M}_1) +
\sum^{\mathrm{N}}_{\n=1}I(\genericRV{M}_2,\genericRVS{Y}^{\n-1}_1,\genericRVS{Y}^{\mathrm{N}}_{2,\n+1},
\genericRVS{W}^{\mathrm{N}}_{\n+1};\genericRV{Y}_{2,n}|\genericRV{M}_1)\\ \nonumber -  \sum^{\mathrm{N}}_{\n=1}I(\genericRV{M}_2;\genericRV{W}_{\n}|\genericRVS{W}^{\mathrm{N}}_{\n+1},
\genericRVS{Y}^{\n-1}_1,\genericRVS{Y}^{\mathrm{N}}_{2,\n+1},\genericRV{M}_1)\\
\nonumber \stackrel{(c)}= \sum^{\mathrm{N}}_{\n=1}I(\genericRVS{Y}^{\n-1}_1;\genericRV{Y}_{2,\n}|\genericRVS{Y}^{\mathrm{N}}_{2,\n+1},\genericRV{M}_1) +
\sum^{\mathrm{N}}_{\n=1}I(\genericRV{M}_2,\genericRVS{Y}^{\n-1}_1,\genericRVS{Y}^{\mathrm{N}}_{2,\n+1},
\genericRVS{W}^{\mathrm{N}}_{\n+1};\genericRV{Y}_{2,\n}|\genericRV{M}_1)\\ -  \sum^{\mathrm{N}}_{\n=1}I(\genericRV{M}_2;\genericRV{W}_{\n}|\genericRVS{W}^{\mathrm{N}}_{\n+1},
\genericRVS{Y}^{\n-1}_1,\genericRVS{Y}^{\mathrm{N}}_{2,\n+1},\genericRV{M}_1) \label{eq:sumrateout2}
\end{eqnarray}
where $(b)$ and $(c)$ follow from Csisz\'{a}r's sum identity. With $\genericRV{U}_{\n}=(\genericRVS{W}^{\mathrm{N}}_{\n+1},\genericRVS{Y}^{\n-1}_1,\genericRVS{Y}^{\mathrm{N}}_{2,\n+1})$; $\genericRV{V}_{1,\n} = \genericRV{M}_1$; and $\genericRV{V}_{2,\n} = \genericRV{M}_2$, from $(\ref{eq:sumrateout1})$ and $(\ref{eq:sumrateout2})$, we get
\begin{eqnarray}
\nonumber \mathrm{N}(R_1+R_2) \leq \sum^{\mathrm{N}}_{n=1}[I(\genericRV{U}_\n, \genericRV{V}_{1,\n};\genericRV{Y}_{1,\n}) - I(\genericRV{V}_{1,\n};\genericRV{W}_\n|\genericRV{U}_\n)]\\  + \sum^{\mathrm{N}}_{\n=1}[I(\genericRV{U}_\n, \genericRV{V}_{2,\n};\genericRV{Y}_{2,\n}|\genericRV{V}_{1,\n}) - I(\genericRV{V}_{2,\n};\genericRV{W}_\n|\genericRV{V}_{1,\n},\genericRV{U}_\n)] + 2\mathrm{N}\epsilon. \label{eq:outboundnewC1R1plusR21}
\end{eqnarray}

Similarly, it can be shown that
\begin{eqnarray}
\nonumber \mathrm{N}(R_1+R_2) \leq \sum^{\mathrm{N}}_{\n=1}[I(\genericRV{U}_\n, \genericRV{V}_{2,\n};\genericRV{Y}_{2,\n}) - I(\genericRV{V}_{2,\n};\genericRV{W}_\n|\genericRV{U}_\n)]\\  + \sum^{\mathrm{N}}_{\n=1}[I(\genericRV{U}_\n, \genericRV{V}_{1,\n};\genericRV{Y}_{1,\n}|\genericRV{V}_{2,\n}) - I(\genericRV{V}_{1,\n};\genericRV{W}_\n|\genericRV{V}_{2,\n},\genericRV{U}_\n)] + 2\mathrm{N}\epsilon. \label{eq:outboundnewC1R1plusR22}
\end{eqnarray}

\subsection{Proof of \thrmref{thm:conversethmC2}}\label{subsec:converseproofC2}
For the channel $\mathrm{C}_2$, $\forall \epsilon >0$ and sufficiently small; and for large $\mathrm{N}$, $R_1$ can be bounded as follows:
\begin{eqnarray}
\nonumber \mathrm{N}R_1 &=& H(\genericRV{M}_1) = I(\genericRV{M}_1;\genericRVS{Y}_1^\mathrm{N})+H(\genericRV{M}_1|\genericRVS{Y}_1^\mathrm{N})\\
\nonumber &\stackrel{(a)}\leq& I(\genericRV{M}_1;\genericRVS{Y}_1^\mathrm{N}) + \mathrm{N}\epsilon
\stackrel{(b)}\leq I(\genericRV{M}_1;\genericRVS{Y}_1^\mathrm{N},\genericRV{M}_2) + \mathrm{N}\epsilon
= I(\genericRV{M}_1;\genericRVS{Y}_1^\mathrm{N}|\genericRV{M}_2) + \mathrm{N}\epsilon \\
\nonumber &\stackrel{(c)}=& \sum^{\mathrm{N}}_{\n=1}[H(\genericRV{Y}_{1,\n}|\genericRVS{Y}^{\n-1}_1,\genericRV{M}_2) - H(\genericRV{Y}_{1,\n}|\genericRVS{Y}^{\n-1}_1,\genericRV{M}_1,\genericRV{M}_2)] + \mathrm{N}\epsilon \\
\nonumber &\stackrel{(d)}\leq& \sum^{\mathrm{N}}_{\n=1}[H(\genericRV{Y}_{1,\n}) - H(\genericRV{Y}_{1,\n}|\genericRVS{Y}^{\n-1}_1,\genericRV{M}_1,\genericRV{M}_2)] + \mathrm{N}\epsilon \\
\nonumber &=& \sum^{\mathrm{N}}_{\n=1}I(\genericRV{M}_1,\genericRV{M}_2,\genericRVS{Y}^{\n-1}_1;\genericRV{Y}_{1,\n}) + \mathrm{N}\epsilon \\
\nonumber &=& \sum^{\mathrm{N}}_{\n=1}[I(\genericRV{M}_1,\genericRV{M}_2,\genericRVS{Y}^{\n-1}_1,\genericRVS{W}^{\mathrm{N}}_{\n+1};\genericRV{Y}_{1,\n}) -I(\genericRVS{W}^{\mathrm{N}}_{\n+1};\genericRV{Y}_{1,\n}|\genericRV{M}_1,\genericRV{M}_2,\genericRVS{Y}^{\n-1}_1)] + \mathrm{N}\epsilon \\
\nonumber &\stackrel{(e)}=& \sum^{\mathrm{N}}_{\n=1}[I(\genericRV{M}_1,\genericRV{M}_2,\genericRVS{Y}^{\n-1}_1,\genericRVS{W}^{\mathrm{N}}_{\n+1};\genericRV{Y}_{1,\n}) -I(\genericRV{M}_1;\genericRV{W}_\n|\genericRV{M}_2,\genericRVS{Y}^{\n-1}_1, \genericRVS{W}^{\mathrm{N}}_{\n+1})] + \mathrm{N}\epsilon \\
\nonumber &\stackrel{(f)}=& \sum^{\mathrm{N}}_{\n=1}[I(\genericRV{M}_1,\genericRV{M}_2,\genericRVS{W}^{\mathrm{N}}_{\n+1};\genericRV{Y}_{1,\n}) -I(\genericRV{M}_1,\genericRV{M}_2,\genericRVS{W}^{\mathrm{N}}_{\n+1};\genericRV{W}_\n|\genericRVS{Y}^{\n-1}_1)] + \mathrm{N}\epsilon \\
&\stackrel{(g)}\leq& \sum^{\mathrm{N}}_{\n=1}[I(\genericRV{M}_1,\genericRV{M}_2,\genericRVS{W}^{\mathrm{N}}_{\n+1};\genericRV{Y}_{1,\n}) -I(\genericRV{M}_1,\genericRV{M}_2,\genericRVS{W}^{\mathrm{N}}_{\n+1};\genericRV{W}_\n) + H(\genericRV{M}_1,\genericRV{M}_2,\genericRVS{W}^{\mathrm{N}}_{\n+1})] + \mathrm{N}\epsilon.
\end{eqnarray}
where $(a)$ follows from Fano's inequality; $(b)$ follows from the data-processing inequality; $(c)$ follows from chain rule; $(d)$ follows from the fact that conditioning reduces entropy; $(e)$ follows from Csisz\'{a}r's sum identity; $(f)$ is due to the memoryless nature of the channel; and $(g)$ is obtained after simple calculations. We let $\genericRV{U}_{n}\triangleq (\genericRV{M}_1,\genericRV{M}_2,\genericRVS{W}^{\mathrm{N}}_{n+1})$ and note that this choice satisfies the Markov chain requirement $\genericRV{U} \rightarrow (\genericRV{X},\genericRV{W}) \rightarrow (\genericRV{Y}_1,\genericRV{Y}_2)$  specified in \secref{sec:statementofresults} for the channel $\mathrm{C}_2$ to get
\begin{eqnarray}
\mathrm{N}R_1 \leq \sum^{\mathrm{N}}_{n=1}[I(\genericRV{U}_{n};\genericRV{Y}_{1,n}) - I(\genericRV{U}_{n};\genericRV{W}_n) + H(\genericRV{U}_{n})] + \mathrm{N}\epsilon. \label{eq:converseproofC2R1}
\end{eqnarray}
By symmetry, we get the following bound on $R_2$:
\begin{eqnarray}
\mathrm{N}R_2 \leq \sum^{\mathrm{N}}_{n=1}[I(\genericRV{U}_{n};\genericRV{Y}_{2,n}) - I(\genericRV{U}_{n};\genericRV{W}_n) + H(\genericRV{U}_{n})] + \mathrm{N}\epsilon. \label{eq:converseproofC2R2}
\end{eqnarray}
We note that the factor $H(\genericRV{U}_{n})$ is independent of the distribution characterizing the channel $\mathrm{C}_2$.

\subsection{Proof of \thrmref{thm:conversethmC31}}\label{subsec:converseproofC31}
For the channel $\mathrm{C}_3$, $\forall \epsilon >0$ and sufficiently small; and for large $\mathrm{N}$, $R_1$ can be bounded as follows:
\begin{eqnarray}
\nonumber \mathrm{N}R_1 &=& H(\genericRV{M}_1) = I(\genericRV{M}_1;\genericRVS{Y}_1^\mathrm{N})+H(\genericRV{M}_1|\genericRVS{Y}_1^\mathrm{N})\\
\nonumber &\stackrel{(a)}\leq& I(\genericRV{M}_1;\genericRVS{Y}_1^\mathrm{N}) + \mathrm{N}\epsilon
\stackrel{(b)}\leq I(\genericRV{M}_1;\genericRVS{Y}_1^\mathrm{N}) - I(\genericRV{M}_1;\genericRVS{Y}_2^\mathrm{N}) + 2\mathrm{N}\epsilon \\
\nonumber &=& \sum^{\mathrm{N}}_{n=1}[I(\genericRV{M}_1;\genericRV{Y}_{1,n}|\genericRVS{Y}^{\mathrm{N}}_{1,n+1}) - I(\genericRV{M}_1;\genericRV{Y}_{2,n}|\genericRVS{Y}^{n-1}_2)] + 2\mathrm{N}\epsilon \\
\nonumber &\stackrel{(c)}=& \sum^{\mathrm{N}}_{n=1}[I(\genericRV{M}_1,\genericRVS{Y}^{n-1}_2;\genericRV{Y}_{1,n}|\genericRVS{Y}^{\mathrm{N}}_{1,n+1}) - I(\genericRV{M}_1,\genericRVS{Y}^{\mathrm{N}}_{1,n+1};\genericRV{Y}_{2,n}|\genericRVS{Y}^{n-1}_2)] + 2\mathrm{N}\epsilon \\
\nonumber &\stackrel{(d)}=& \sum^{\mathrm{N}}_{n=1}[I(\genericRV{M}_1;\genericRV{Y}_{1,n}|\genericRVS{Y}^{\mathrm{N}}_{1,n+1},\genericRVS{Y}^{n-1}_2) - I(\genericRV{M}_1;\genericRV{Y}_{2,n}|\genericRVS{Y}^{\mathrm{N}}_{1,n+1},\genericRVS{Y}^{n-1}_2)] + 2\mathrm{N}\epsilon \\
\nonumber &\leq& \sum^{\mathrm{N}}_{n=1}[I(\genericRV{M}_1,\genericRV{W}_n;\genericRV{Y}_{1,n}|\genericRVS{Y}^{\mathrm{N}}_{1,n+1},\genericRVS{Y}^{n-1}_2) - I(\genericRV{M}_1;\genericRV{Y}_{2,n}|\genericRVS{Y}^{\mathrm{N}}_{1,n+1},\genericRVS{Y}^{n-1}_2)] + 2\mathrm{N}\epsilon \\
\nonumber &\stackrel{(e)}=& \sum^{\mathrm{N}}_{n=1}[I(\genericRV{M}_1;\genericRV{Y}_{1,n}|\genericRVS{Y}^{\mathrm{N}}_{1,n+1},\genericRVS{Y}^{n-1}_2)+ I(\genericRV{W}_n;\genericRV{Y}_{1,n}|\genericRV{M}_1,\genericRVS{Y}^{\mathrm{N}}_{1,n+1},\genericRVS{Y}^{n-1}_2)\\
\nonumber &&-I(\genericRV{M}_1;\genericRV{Y}_{2,n}|\genericRVS{Y}^{\mathrm{N}}_{1,n+1},\genericRVS{Y}^{n-1}_2)] + 2\mathrm{N}\epsilon \\
\nonumber &=& \sum^{\mathrm{N}}_{n=1}[I(\genericRV{M}_1;\genericRV{Y}_{1,n}|\genericRVS{Y}^{\mathrm{N}}_{1,n+1},\genericRVS{Y}^{n-1}_2) + H(\genericRV{W}_n|\genericRV{M}_1,\genericRVS{Y}^{\mathrm{N}}_{1,n+1},\genericRVS{Y}^{n-1}_2)\\
\nonumber &&- H(\genericRV{W}_n|\genericRV{M}_1,\genericRV{Y}_{1,n},\genericRVS{Y}^{\mathrm{N}}_{1,n+1},\genericRVS{Y}^{n-1}_2) - I(\genericRV{M}_1;\genericRV{Y}_{2,n}|\genericRVS{Y}^{\mathrm{N}}_{1,n+1},\genericRVS{Y}^{n-1}_2)] + 2\mathrm{N}\epsilon \\
\nonumber &\leq& \sum^{\mathrm{N}}_{n=1}[I(\genericRV{M}_1;\genericRV{Y}_{1,n}|\genericRVS{Y}^{\mathrm{N}}_{1,n+1},\genericRVS{Y}^{n-1}_2) + H(\genericRV{W}_n|\genericRV{M}_1,\genericRVS{Y}^{\mathrm{N}}_{1,n+1},\genericRVS{Y}^{n-1}_2)\\
\nonumber &&-I(\genericRV{M}_1;\genericRV{Y}_{2,n}|\genericRVS{Y}^{\mathrm{N}}_{1,n+1},\genericRVS{Y}^{n-1}_2)] + 2\mathrm{N}\epsilon,
\end{eqnarray}
where $(a)$ is from Fano's inequality, $(b)$ is from confidentiality constraints, $(c)$ and $(d)$ follow from Csisz\'{a}r's sum identity and $(e)$ is the chain rule for mutual information. Letting $\genericRV{U}_n \triangleq (\genericRVS{Y}^{\mathrm{N}}_{1,n+1},\genericRVS{Y}^{n-1}_2)$; and $\genericRV{V}_{1,1} = \dots = \genericRV{V}_{1,\mathrm{N}} \triangleq \genericRV{M}_1$, where $\genericRV{U}$ and $\genericRV{V}_{1}$ satisfy the Markov chain $\genericRV{U}\rightarrow \genericRV{V}_1 \rightarrow \genericRV{X}$ specified in \secref{sec:statementofresults} for the channel $\mathrm{C}_3$, we get
\begin{eqnarray}
\mathrm{N}R_1 &\leq& \sum^{\mathrm{N}}_{n=1}[I(\genericRV{V}_{1,n};\genericRV{Y}_{1,n}|\genericRV{U}_n)+H(\genericRV{W}_n|\genericRV{U}_n,\genericRV{V}_{1,n}) - I(\genericRV{V}_{1,n};\genericRV{Y}_{2,n}|\genericRV{U}_n)] + 2\mathrm{N}\epsilon. \label{eq:converseproofC3R1}
\end{eqnarray}
Proceeding in a similar fashion and letting $\genericRV{V}_{2,1} = \dots = \genericRV{V}_{2,\mathrm{N}} \triangleq \genericRV{M}_2$,
\begin{eqnarray}
\mathrm{N}R_2 &\leq& \sum^{\mathrm{N}}_{n=1}[I(\genericRV{V}_{2,n};\genericRV{Y}_{2,n}|\genericRV{U}_n)+H(\genericRV{W}_n|\genericRV{U}_n,\genericRV{V}_{2,n}) - I(\genericRV{V}_{2,n};\genericRV{Y}_{1,n}|U_n)] + 2\mathrm{N}\epsilon. \label{eq:converseproofC3R2}
\end{eqnarray}

For the channel $\mathrm{C}_3$, we also derive a genie-aided outer bound by letting a hypothetical genie give $\mathrm{D}_1$ message $\genericRV{M}_2$, while $\mathrm{D}_2$ computes the equivocation using $\genericRV{M}_2$ as side-information. $\forall \epsilon >0$ and sufficiently small; and for large $\mathrm{N}$, $R_1$ can be upper bounded as follows:
\begin{eqnarray}
\nonumber \mathrm{N}R_1 &=& H(\genericRV{M}_1) \leq H(\genericRV{M}_1|\genericRVS{Y}_2^\mathrm{N}) + \mathrm{N}\epsilon
\leq H(\genericRV{M}_1,\genericRV{M}_2|\genericRVS{Y}_2^\mathrm{N}) + \mathrm{N}\epsilon\\
\nonumber &=& H(\genericRV{M}_1|\genericRVS{Y}_2^\mathrm{N},\genericRV{M}_2) + H(\genericRV{M}_2|\genericRVS{Y}_2^\mathrm{N})  + \mathrm{N}\epsilon
\leq H(\genericRV{M}_1|\genericRVS{Y}_2^\mathrm{N},\genericRV{M}_2) + \mathrm{N}\epsilon\\
\nonumber &\leq& H(\genericRV{M}_1|\genericRVS{Y}_2^\mathrm{N},\genericRV{M}_2) - H(\genericRV{M}_1|\genericRVS{Y}_1^\mathrm{N}) + \mathrm{N}\epsilon
\stackrel{(a)}\leq H(\genericRV{M}_1|\genericRVS{Y}_2^\mathrm{N},\genericRV{M}_2) - H(\genericRV{M}_1|\genericRVS{Y}_1^\mathrm{N},\genericRV{M}_2) + \mathrm{N}\epsilon\\
\nonumber &\leq& I(\genericRV{M}_1;\genericRVS{Y}_1^\mathrm{N}|\genericRV{M}_2) - I(\genericRV{M}_1;\genericRVS{Y}_2^\mathrm{N}|\genericRV{M}_2) + 2\mathrm{N}\epsilon \\
\nonumber &=& \sum^{\mathrm{N}}_{n=1}[I(\genericRV{M}_1;\genericRV{Y}_{1,n}|\genericRVS{Y}^{\mathrm{N}}_{1,n+1},\genericRV{M}_2) - I(\genericRV{M}_1;\genericRV{Y}_{2,n}|\genericRVS{Y}^{n-1}_2,\genericRV{M}_2)] + 2\mathrm{N}\epsilon \\
\nonumber &\stackrel{(b)}=& \sum^{\mathrm{N}}_{n=1}[I(\genericRV{M}_1,\genericRVS{Y}^{n-1}_2;\genericRV{Y}_{1,n}|\genericRVS{Y}^{\mathrm{N}}_{1,n+1},\genericRV{M}_2) - I(\genericRV{M}_1,\genericRVS{Y}^{\mathrm{N}}_{1,n+1};\genericRV{Y}_{2,n}|\genericRVS{Y}^{n-1}_2,\genericRV{M}_2)] + 2\mathrm{N}\epsilon \\
\nonumber &\stackrel{(c)}=& \sum^{\mathrm{N}}_{n=1}[I(\genericRV{M}_1;\genericRV{Y}_{1,n}|\genericRVS{Y}^{\mathrm{N}}_{1,n+1},\genericRVS{Y}^{n-1}_2,\genericRV{M}_2) - I(\genericRV{M}_1;\genericRV{Y}_{2,n}|\genericRVS{Y}^{\mathrm{N}}_{1,n+1},\genericRVS{Y}^{n-1}_2,\genericRV{M}_2)] + 2\mathrm{N}\epsilon \\
\nonumber &\leq& \sum^{\mathrm{N}}_{n=1}[I(\genericRV{M}_1,\genericRV{W}_n;\genericRV{Y}_{1,n}|\genericRVS{Y}^{\mathrm{N}}_{1,n+1},\genericRVS{Y}^{n-1}_2,\genericRV{M}_2) - I(\genericRV{M}_1;\genericRV{Y}_{2,n}|\genericRVS{Y}^{\mathrm{N}}_{1,n+1},\genericRVS{Y}^{n-1}_2,\genericRV{M}_2)] + 2\mathrm{N}\epsilon \\
\nonumber &=& \sum^{\mathrm{N}}_{n=1}[I(\genericRV{M}_1;\genericRV{Y}_{1,n}|\genericRVS{Y}^{\mathrm{N}}_{1,n+1},\genericRVS{Y}^{n-1}_2,\genericRV{M}_2) + I(\genericRV{W}_n;\genericRV{Y}_{1,n}|\genericRV{M}_1,\genericRVS{Y}^{\mathrm{N}}_{1,n+1},\genericRVS{Y}^{n-1}_2,\genericRV{M}_2)\\
\nonumber &&-I(\genericRV{M}_1;\genericRV{Y}_{2,n}|\genericRVS{Y}^{\mathrm{N}}_{1,n+1},\genericRVS{Y}^{n-1}_2,\genericRV{M}_2)] + 2\mathrm{N}\epsilon \\
\nonumber &=& \sum^{\mathrm{N}}_{n=1}[I(\genericRV{M}_1;\genericRV{Y}_{1,n}|\genericRVS{Y}^{\mathrm{N}}_{1,n+1},\genericRVS{Y}^{n-1}_2,\genericRV{M}_2) + H(\genericRV{W}_n|\genericRV{M}_1,Y^{\mathrm{N}}_{n+1},\genericRVS{Y}^{n-1}_2,\genericRV{M}_2)\\
\nonumber &&- H(\genericRV{W}_n|\genericRV{M}_1,\genericRV{Y}_{1,n},\genericRVS{Y}^{\mathrm{N}}_{1,n+1},\genericRVS{Y}^{n-1}_2,\genericRV{M}_2) - I(\genericRV{M}_1;\genericRV{Y}_{2,n}|\genericRVS{Y}^{\mathrm{N}}_{1,n+1},\genericRVS{Y}^{n-1}_2,\genericRV{M}_2)] + 2\mathrm{N}\epsilon \\
\nonumber &\leq& \sum^{\mathrm{N}}_{n=1}[I(\genericRV{M}_1;\genericRV{Y}_{1,n}|\genericRVS{Y}^{\mathrm{N}}_{1,n+1},\genericRVS{Y}^{n-1}_2,\genericRV{M}_2) + H(\genericRV{W}_n|\genericRV{M}_1,\genericRVS{Y}^{\mathrm{N}}_{1,n+1},\genericRVS{Y}^{n-1}_2,\genericRV{M}_2)\\
\nonumber &&-I(\genericRV{M}_1;\genericRV{Y}_{2,n}|\genericRVS{Y}^{\mathrm{N}}_{1,n+1},\genericRVS{Y}^{n-1}_2,\genericRV{M}_2)] + 2\mathrm{N}\epsilon,
\end{eqnarray}
where $(a)$ follows since the genie gives $\mathrm{D}_1$ message $\genericRV{M}_2$, $(b)$ and $(c)$ follow from Csisz\'{a}r's sum identity. Letting $\genericRV{U}_n \triangleq (\genericRVS{Y}^{\mathrm{N}}_{1,n+1},\genericRVS{Y}^{n-1}_2)$, $\genericRV{V}_{1,1} = \dots = \genericRV{V}_{1,\mathrm{N}} \triangleq \genericRV{M}_1$ and $\genericRV{V}_{2,1} = \dots = \genericRV{V}_{2,\mathrm{N}} \triangleq \genericRV{M}_2$, where $\genericRV{U}$, $\genericRV{V}_{1}$ and $\genericRV{V}_{2}$ satisfy the Markov chains $\genericRV{U}\rightarrow \genericRV{V}_1 \rightarrow \genericRV{X}$ and $\genericRV{U}\rightarrow \genericRV{V}_2 \rightarrow \genericRV{X}$ specified in \secref{sec:statementofresults} for the channel $\mathrm{C}_3$, $R_1$ can be bounded as
\begin{eqnarray}
\mathrm{N}R_1 \leq \sum^{\mathrm{N}}_{n=1}[I(\genericRV{V}_{1,n};\genericRV{Y}_{1,n}|\genericRV{U}_n,\genericRV{V}_{2,n}) + H(\genericRV{W}_n|\genericRV{U}_n,\genericRV{V}_{1,n},\genericRV{V}_{2,n})
- I(\genericRV{V}_{1,n};\genericRV{Y}_{2,n}|\genericRV{U}_n,\genericRV{V}_{2,n})] + 2\mathrm{N}\epsilon. \label{eq:converseproofC3R1genie}
\end{eqnarray}
Similarly,
\begin{eqnarray}
\mathrm{N}R_1 \leq \sum^{\mathrm{N}}_{n=1}[I(\genericRV{V}_{2,n};\genericRV{Y}_{2,n}|\genericRV{U}_n,\genericRV{V}_{1,n}) + H(\genericRV{W}_n|\genericRV{U}_n,\genericRV{V}_{1,n},\genericRV{V}_{2,n})
- I(\genericRV{V}_{2,n};\genericRV{Y}_{1,n}|\genericRV{U}_n,\genericRV{V}_{1,n})] + 2\mathrm{N}\epsilon.  \label{eq:converseproofC3R2genie}
\end{eqnarray}

For the channel $\mathrm{C}_3$, the outer bound on $R_1+R_2$ can be made tighter by the following procedure. From $(\ref{eq:minoutboundC3R1})$ - $(\ref{eq:minoutboundC3R2})$, we see that
\begin{eqnarray}
R_1+R_2 \leq I_1 + I_2, \label{eq:sumratebound1}\\
R_1+R_2 \leq I^{\ast}_1 + I^{\ast}_2. \label{eq:sumratebound2}
\end{eqnarray}
Therefore,
\begin{eqnarray}
R_1+R_2 \leq \min [I_1 + I^{\ast}_2, I_2 + I^{\ast}_1]. \label{eq:sumratebound}
\end{eqnarray}
We show now that the bound $(\ref{eq:sumratebound})$ is a tighter bound than $(\ref{eq:sumratebound1})$ and $(\ref{eq:sumratebound2})$. It is easy to see that
\begin{eqnarray}
\nonumber I_1 + I_2 = I^{\ast}_1 + I^{\ast}_2 + I(W;V_1|U,V_2) + I(W;V_2|U,V_1).
\end{eqnarray}
Consider $2(I_1 + I_2) = 2[I^{\ast}_1 + I^{\ast}_2 + I(W;V_1|U,V_2) + I(W;V_2|U,V_1)]$, which implies the following:
\begin{eqnarray}
\nonumber \min[I_1 + I^{\ast}_2, I_2 + I^{\ast}_1] &\leq& I_1 + I_2,\\
\nonumber \min[I_1 + I^{\ast}_2, I_2 + I^{\ast}_1] &\leq& I^{\ast}_1 + I^{\ast}_2.
\end{eqnarray}
Therefore, the sum rate bound given by $(\ref{eq:sumratebound})$ is tighter than $(\ref{eq:sumratebound1})$ and $(\ref{eq:sumratebound2})$.

\subsection{Proof of \thrmref{thm:conversethmC32}}\label{subsec:converseproofC32}
For the channel $\mathrm{C}_3$, $\forall \epsilon >0$ and sufficiently small; and for large $\mathrm{N}$, $R_1$ can be bounded as follows:
\begin{eqnarray}
\nonumber \mathrm{N}R_1 &=& H(\genericRV{M}_1) = I(\genericRV{M}_1;\genericRVS{Y}_1^\mathrm{N})+H(\genericRV{M}_1|\genericRVS{Y}_1^\mathrm{N})\\
\nonumber &\stackrel{(a)}\leq& I(\genericRV{M}_1;\genericRVS{Y}_1^\mathrm{N}) + \mathrm{N}\epsilon
\stackrel{(b)}\leq I(\genericRV{M}_1;\genericRVS{Y}_1^\mathrm{N}) - I(\genericRV{M}_1;\genericRVS{Y}_2^\mathrm{N}) + 2\mathrm{N}\epsilon,
\end{eqnarray}
where $(a)$ follows from Fano's inequality; and $(b)$ follows from confidentiality constraints. Following the procedure used to prove \thrmref{thm:conversethmC12} (see \secref{subsec:converseproofC12}) and letting $\genericRV{U}_{\mathrm{n}}=(\genericRVS{W}^{\mathrm{N}}_{\mathrm{n}+1},\genericRVS{Y}^{\mathrm{n}-1}_1,\genericRVS{Y}^{\mathrm{N}}_{2,\mathrm{n}+1})$ and $\genericRV{V}_{1,\mathrm{n}} = \genericRV{M}_1$,
\begin{eqnarray}
\mathrm{N}R_1 \leq \sum^{\mathrm{N}}_{\mathrm{n}=1}I(\genericRV{U}_{\mathrm{n}}, \genericRV{V}_{1,\mathrm{n}};\genericRV{Y}_{1,\mathrm{n}}) - I(\genericRV{V}_{1,\mathrm{n}};\genericRV{W}_{\mathrm{n}}|\genericRV{U}_{\mathrm{n}}) -
I(\genericRV{V}_{1,\mathrm{n}};\genericRV{Y}_{2,\mathrm{n}}) + 2\mathrm{N}\epsilon.
\end{eqnarray}
Similarly, letting $\genericRV{V}_{2,\mathrm{n}} = \genericRV{M}_2$, we get
\begin{eqnarray}
\mathrm{N}R_2 \leq \sum^{\mathrm{N}}_{\mathrm{n}=1}I(\genericRV{U}_{\mathrm{n}}, \genericRV{V}_{2,\mathrm{n}};\genericRV{Y}_{2,\mathrm{n}}) - I(\genericRV{V}_{2,\mathrm{n}};\genericRV{W}_{\mathrm{n}}|\genericRV{U}_{\mathrm{n}}) - I(\genericRV{V}_{2,\mathrm{n}};\genericRV{Y}_{1,\mathrm{n}}) + 2\mathrm{N}\epsilon,
\end{eqnarray}
and the following bounds on the sum-rate $R_1+R_2$:
\begin{eqnarray}
\nonumber \mathrm{N}(R_1+R_2) \leq \sum^{\mathrm{N}}_{n=1}[I(\genericRV{U}_\n, \genericRV{V}_{1,\n};\genericRV{Y}_{1,\n}) - I(\genericRV{V}_{1,\n};\genericRV{W}_\n|\genericRV{U}_\n)]\\ + \sum^{\mathrm{N}}_{\n=1}[I(\genericRV{U}_\n, \genericRV{V}_{2,\n};\genericRV{Y}_{2,\n}|\genericRV{V}_{1,\n}) - I(\genericRV{V}_{2,\n};\genericRV{W}_\n|\genericRV{V}_{1,\n},\genericRV{U}_\n)] - I(\genericRV{V}_{1,\n};\genericRV{Y}_{2,\n}) + 2\mathrm{N}\epsilon,\\
\nonumber \mathrm{N}(R_1+R_2) \leq \sum^{\mathrm{N}}_{\n=1}[I(\genericRV{U}_\n, \genericRV{V}_{2,\n};\genericRV{Y}_{2,\n}) - I(\genericRV{V}_{2,\n};\genericRV{W}_\n|\genericRV{U}_\n)]\\  + \sum^{\mathrm{N}}_{\n=1}[I(\genericRV{U}_\n, \genericRV{V}_{1,\n};\genericRV{Y}_{1,\n}|\genericRV{V}_{2,\n}) - I(\genericRV{V}_{1,\n};\genericRV{W}_\n|\genericRV{V}_{2,\n},\genericRV{U}_\n)] - I(\genericRV{V}_{2,\n};\genericRV{Y}_{1,\n}) +  2\mathrm{N}\epsilon.
\end{eqnarray}

A time sharing RV $\genericRV{Q}$, which is uniformly distributed over $\mathrm{N}$ symbols and independent of the RVs $\genericRV{M}_1$, $\genericRV{M}_2$, $\genericRV{W}$, $\genericRV{U}$, $\genericRV{V}_1$, $\genericRV{V}_2$, $\genericRV{X}$, $\genericRV{Y}_1$ and $\genericRV{Y}_2$ is introduced for the single letter characterization of the above derived outer bounds. Applying the procedure similar to the one presented in \cite[Chapter 15.3.4]{Cover2006} on the $\mathrm{N}$-letter expressions obtained in the above stated theorems, we get the outer bounds presented in \secref{sec:statementofresults}. This completes the proofs of
\thrmref{thm:conversethmC1}, \thrmref{thm:conversethmC12}, \thrmref{thm:conversethmC2}, \thrmref{thm:conversethmC31} and \thrmref{thm:conversethmC32}.

\section{Conclusions}\label{sec:conclusions}
We presented inner and outer bounds on the capacity region of three classes of two-user discrete memoryless broadcast channels, with noncausal side-information at the encoder. We generalized existing approaches to prove the achievability theorems, and characterized the rate penalties for having to deal with side-information at the encoder. For channels with confidentiality constraints, we showed that rate penalties exist for dealing with both side-information and confidentiality constraints. In the case of outer bounds, we focus on the explicit characterization of the sum-rate bounds. For channels where each receiver has {\apriori} knowledge of the message of the other receiver, we showed that the outer bounds are only a factor away from the achievable region, where the factor is independent of the channel distribution. 

\appendices
\section{Encoder error analysis}\label{appendix:errorencodeC1}
Here, we upper bound the probability of encoder error for the channel $\mathrm{C}_1$, by using results from the second moment method \cite{Alon2000}. This method was also employed in \cite{Gamal1981} and \cite[Chap. 7, pp. 354]{Kramer2007a} to provide an alternative proof of Marton's achievability scheme. An error is declared at the encoder of $\mathrm{S}$ if it is not possible to find a pair $(i_1,i_2)$ to satisfy the condition $E_{\mathrm{S}} \triangleq \{(\genericRVS{W},\genericRVS{V}_1(i_1,j_1),\genericRVS{V}_2(i_2,j_2))\in A^{(\mathrm{N})}_{\epsilon}(P_{\genericRV{W},\genericRV{V}_1,\genericRV{V}_2})\}$. Let $P_{e,E_{\mathrm{S}}}$ denote the probability of error at the encoder, \ie, $P_{e,E_{\mathrm{S}}} \triangleq \text{Pr}(E_{\mathrm{S}}^c)$. Let $\genericRV{I}$ be an indicator RV that the event $E_{\mathrm{S}}$ has occurred. Let $\genericRV{Q} = \sum_{j_1,j_2} \genericRV{I}$; $\bar{\genericRV{Q}} = \mathbb{E}[\genericRV{Q}]$; and $\text{Var}[\genericRV{Q}] = \mathbb{E}[(\genericRV{Q}-\bar{\genericRV{Q}})^2]$, where $\mathbb{E}(.)$ denotes the expectation operator. $P_{e,E_{\mathrm{S}}}$ can be upper bounded as follows:
\begin{eqnarray}
P_{e,E_{\mathrm{S}}} = \text{Pr}(\genericRV{Q}=0) \stackrel{(i)}{\leq} \text{Var}[\genericRV{Q}]/\bar{\genericRV{Q}}^2, \label{eq:appenderrorencodeC1PeS}
\end{eqnarray}
where $(i)$ follows from Markov's inequality for non-negative RVs. Consider now
\begin{eqnarray*}
\bar{\genericRV{Q}} &=& \sum_{j_1,j_2}\mathbb{E}(\genericRV{I}) \geq \sum_{j_1,j_2}(1-\delta^{(\mathrm{N})})2^{-\mathrm{N}[I(\genericRV{V}_1;\genericRV{V}_2)+I(\genericRV{V}_1,\genericRV{V}_2;\genericRV{W})
+4\epsilon]}\\
&=& (1-\delta^{(\mathrm{N})})2^{-\mathrm{N}[R^{\ast}_1+R^{\ast}_2 - I(\genericRV{V}_1;\genericRV{V}_2) - I(\genericRV{V}_1,\genericRV{V}_2;\genericRV{W}) - 4\epsilon]}.
\end{eqnarray*}
Next, consider $\text{Var}[\genericRV{Q}] = \sum_{j_1,j_2}\sum_{j'_1,j'_2}\{\mathbb{E}[\genericRV{I}(j_1,j_2)\genericRV{I}(j'_1,j'_2)] - \mathbb{E}[\genericRV{I}(j_1,j_2)]\mathbb{E}\genericRV{I}(j'_1,j'_2)]\}$. We have the following four cases:
\begin{enumerate}
\item If $j'_1\neq j_1$ and $j'_2\neq j_2$, then $\genericRV{I}(j_1,j_2)$ and $\genericRV{I}(j'_1,j'_2)$ are independent and $\text{Var}[\genericRV{Q}]=0$.
\item If $j'_1 = j_1$ and $j'_2 = j_2$, then $\mathbb{E}[\genericRV{I}(j_1,j_2)\genericRV{I}(j'_1,j'_2)] = \mathbb{E}[\genericRV{I}(j_1,j_2)]\leq 2^{-\mathrm{N}[I(\genericRV{V}_1;\genericRV{V}_2)+I(\genericRV{V}_1,\genericRV{V}_2;\genericRV{W})-4\epsilon]}$.
\item If $j'_1 \neq j_1$ and $j'_2 = j_2$, then $\mathbb{E}[\genericRV{I}(j_1,j_2)\genericRV{I}(j'_1,j'_2)] \leq 2^{-\mathrm{N}[I(\genericRV{V}_1;\genericRV{V}_2|U)+I(\genericRV{V}_1,\genericRV{V}_2;\genericRV{W})+I(\genericRV{V}_1;\genericRV{V}_2,\genericRV{W})-6\epsilon]}$.
\item If $j'_1 = j_1$ and $j'_2 \neq j_2$, then $\mathbb{E}[\genericRV{I}(j_1,j_2)\genericRV{I}(j'_1,j'_2)] \leq 2^{-\mathrm{N}[I(\genericRV{V}_1;\genericRV{V}_2|U)+I(\genericRV{V}_1,\genericRV{V}_2;\genericRV{W})+I(\genericRV{V}_2;\genericRV{V}_1,\genericRV{W})-6\epsilon]}$.
\end{enumerate}

Substituting for $\bar{\genericRV{Q}}$ and $\text{Var}[\genericRV{Q}]$ in $(\ref{eq:appenderrorencodeC1PeS})$, we can show that $P(E_{\mathrm{S}}) \leq \delta^{(\mathrm{N})}_{\mathrm{C}_1}$, $\forall \delta^{(\mathrm{N})}_{\mathrm{C}_1} > 0$ and sufficiently small; and for $\mathrm{N}$ large, if the following conditions are simultaneously satisfied:
\begin{eqnarray}
R_1^{'} &>& I(\genericRV{W};\genericRV{V}_1) - \epsilon_1,\label{eq:appenderrorencodeC1R1'}\\
R_2^{'} &>& I(\genericRV{W};\genericRV{V}_2) - \epsilon_2,\label{eq:appenderrorencodeC1R2'}\\
R_1^{'} + R_2^{'} &>& I(\genericRV{V}_1;\genericRV{V}_2) + I(\genericRV{V}_1,\genericRV{V}_2;\genericRV{W}) - \epsilon_3. \label{eq:appenderrorencodeC1R1plusR2'}
\end{eqnarray}

Similar analysis results in a bound on the binning rates for the channel $\mathrm{C}_3$. The probability of encoder error $P(E_{\mathrm{S}}) \leq \delta^{(\mathrm{N})}_{\mathrm{C}_3}$, $\forall \delta^{(\mathrm{N})}_{\mathrm{C}_3} > 0$ and sufficiently small; and for $\mathrm{N}$ large, if the following conditions are simultaneously satisfied:
\begin{eqnarray}
R_1^{\ast} &>& I(\genericRV{W};\genericRV{V}_1|U) - \epsilon_{11},\label{eq:appenderrorencodeC3R1*}\\
R_2^{\ast} &>& I(\genericRV{W};\genericRV{V}_2|U) - \epsilon_{12},\label{eq:appenderrorencodeC3R2*}\\
R_1^{\ast} + R_2^{\ast} &>& I(\genericRV{V}_1;\genericRV{V}_2|U) + I(\genericRV{V}_1,\genericRV{V}_2;\genericRV{W}|U) - \epsilon_{13}. \label{eq:appenderrorencodeC3R1plusR2*}
\end{eqnarray}


\bibliographystyle{IEEEtran}
\bibliography{IEEEabrv,bc_journal}
\clearpage
\end{document}